\documentclass[10pt]{article}
\usepackage[numbers,sort&compress]{natbib}
\usepackage{amsmath}
\usepackage{amssymb}
\usepackage{graphicx}
\usepackage{color}
\usepackage[table,xcdraw]{xcolor}
\usepackage{etoolbox}   
\usepackage{orcidlink}
\usepackage{hyperref}
\hypersetup{colorlinks=true, linkcolor=blue, citecolor=blue, urlcolor=blue}
\usepackage{array}
\usepackage{longtable}
\usepackage{booktabs}
\usepackage{colortbl}
\usepackage{subcaption}
\usepackage{geometry}
\usepackage{colortbl,xcolor}
\usepackage{tikz}
\usepackage{array}
\usepackage{diagbox}
\renewcommand{\arraystretch}{1.8}

\RequirePackage[numbers,sort&compress]{natbib}
\begin{document}
\baselineskip=20pt

\begin{center}
{\LARGE{\bf Observational signatures of charged Bardeen black holes in perfect fluid dark matter with a cloud of strings}}
\end{center}

\begin{center}
{\bf Faizuddin Ahmed}\orcidlink{0000-0003-2196-9622}\\
Department of Physics, The Assam Royal Global University, Guwahati, 781035, Assam, India\\
e-mail: faizuddinahmed15@gmail.com\\

\vspace{0.1cm}
{\bf Ahmad Al-Badawi}\orcidlink{0000-0002-3127-3453}\\
Department of Physics, Al-Hussein Bin Talal University, 71111, Ma'an, Jordan. \\
e-mail: ahmadbadawi@ahu.edu.jo\\

\vspace{0.1cm}
{\bf \.{I}zzet Sakall{\i}}\orcidlink{0000-0001-7827-9476}\\
Physics Department, Eastern Mediterranean University, Famagusta 99628, North Cyprus via Mersin 10, Turkey\\
e-mail: izzet.sakalli@emu.edu.tr (Corresponding author)
\end{center}

\begin{abstract}
We construct a charged Bardeen black hole (BH) surrounded by perfect fluid dark matter (PFDM) and coupled to a cloud of strings (CS). The metric function combines the magnetic monopole charge from nonlinear electrodynamics, the PFDM logarithmic correction, and the CS parameter that renders the spacetime asymptotically non-flat. We analyze the horizon structure, identifying parameter ranges yielding non-extremal BHs, extremal configurations, and naked singularities. The null geodesics, photon sphere radius, and shadow are computed, revealing that both CS and PFDM enlarge the shadow. For neutral particle dynamics, we derive the specific energy, angular momentum, and innermost stable circular orbit location. Quasiperiodic oscillations (QPOs) are examined through the azimuthal, radial, and vertical epicyclic frequencies, where notably the azimuthal frequency is independent of the CS parameter. Scalar field perturbations governed by the Klein-Gordon equation yield an effective potential whose peak decreases with both parameters, yet the transmission and reflection probabilities respond oppositely to CS and PFDM variations. The greybody factor bounds are obtained using semi-analytical methods. Our results demonstrate that the distinct effects of $\alpha$ and $\beta$ on various observables could allow independent constraints on these parameters through shadow measurements, QPO timing, and gravitational wave ringdown observations.\\

\end{abstract}

\tableofcontents

{\color{black}
\section{Introduction} \label{isec1}

The existence of BHs, predicted by Einstein's general relativity (GR), has been confirmed through multiple observational channels including gravitational wave (GW) detections by the LIGO-Virgo-KAGRA collaboration \cite{isz01,isz02} and direct imaging of supermassive BH shadows by the Event Horizon Telescope (EHT) collaboration \cite{isz03,isz03b,isz03c,isz04,isz04b,isz04c}. These observations have opened new windows for testing gravity in the strong-field regime and probing the nature of compact objects. However, classical BH solutions such as Schwarzschild and Reissner-Nordström (RN) suffer from the central curvature singularity where geodesic incompleteness occurs and physical quantities diverge \cite{isz05,isz06}. This singularity problem has motivated the search for regular BH solutions that maintain finite curvature throughout the spacetime.

The first regular BH solution was proposed by Bardeen in 1968 \cite{isz07}, which avoids the central singularity by introducing a modification to the Schwarzschild metric. Later, Ayón-Beato and García demonstrated that the Bardeen solution can be interpreted as a gravitational field of a nonlinear magnetic monopole arising from nonlinear electrodynamics (NED) coupled to GR \cite{isz08,isz09}. In this framework, the magnetic charge $q$ regularizes the geometry near $r = 0$, replacing the singularity with a de Sitter-like core where the spacetime curvature remains finite \cite{isz10,isz11}. The Bardeen BH has since become a prototype for regular BH studies, with extensions including rotation \cite{isz12}, additional charges \cite{isz13}, and coupling to various matter fields \cite{isz14,isz15,AHMED2025101907,ALBADAWI2020168026,Sucu:2025fwa,Aydiner:2025eii,Gursel:2025wan,
Sucu:2025pce,Sakalli:2025els,Sucu:2025eix}.

Astrophysical BHs are not isolated objects but exist within environments containing various forms of matter and energy. Dark matter, which constitutes approximately 27\% of the total energy density of the universe, is expected to accumulate around massive compact objects due to gravitational attraction \cite{isz16,isz17}. Among the phenomenological models describing dark matter distributions around BHs, PFDM has attracted considerable attention \cite{isz18,isz19}. Unlike the quintessence model characterized by the equation of state parameter $\omega_q$, PFDM introduces an anisotropic pressure profile with $P_r \neq P_\theta = P_\phi$, leading to logarithmic corrections in the metric function of the form $(\beta/r)\ln(r/\beta)$, where $\beta > 0$ quantifies the dark matter intensity \cite{isz20,isz21}. Studies have shown that PFDM modifies the horizon structure, photon sphere (PS), shadow, and thermodynamic properties of BHs \cite{isz22,isz23}. Various authors have studied PFDM as an external matter field in different black hole configurations-both within Einstein’s theory and in modified gravity theories frameworks-and have investigated their thermodynamic properties, quasinormal modes, shadows, and topological classes, analyzing the resulting outcomes (see Refs. \cite{Xu2018_Kerr_PFDM,Rizwan2019_PFDM_Kerr,Hendi2020_PFDM,Rizwan2023_PFDM,Liang2023_PFDM,Rizwan2025_PFDM,Sekhmani2025_PFDM_ModMax,QT2025,Ali2025_PFDM,Shi2024_PFDM_EulerHeisenberg}).  

Another astrophysically motivated source that can modify BH geometry is the cloud of strings (CS), introduced by Letelier \cite{isz24}. Cosmic strings are one-dimensional topological defects that may have formed during phase transitions in the early universe \cite{isz25,isz26}. A CS represents a collection of strings extending radially from the central object, and its gravitational effect enters the metric through the parameter $\alpha \in [0,1)$, which shifts the asymptotic value of the metric function to $f(r \to \infty) = 1 - \alpha$ \cite{isz27,isz28}. The CS renders the spacetime asymptotically non-flat, affecting the PS radius, shadow observables, and particle dynamics \cite{isz29,isz30}. Researchers have investigated both rotating and non-rotating black holes in Einstein as well as modified gravity theories, considering the presence of a cloud of strings and surrounding matter fields (see Refs.~\cite{Yang2024_String, Nascimento2024_String, Atamurotov2023_String, Ahmed2026_NPB_String, Mustafa2022_String, Molla2024_String, Ahmed2025_String, AlBadawi2024_String, Waseem2023_String, Nascimento2024_AOP_String}).

The PS is the surface of unstable circular photon orbits surrounding a BH, and its radius $r_{\rm ph}$ directly determines the BH shadow size as observed by distant observers \cite{isz31}. For asymptotically flat spacetimes, the shadow radius equals the critical impact parameter, while for non-flat cases such as those with CS, appropriate corrections must be applied \cite{isz32}. The EHT observations of M87* and Sgr A* have provided shadow measurements that can be used to constrain deviations from the Kerr metric, making shadow studies a powerful tool for testing alternative gravity theories and exotic matter distributions \cite{isz03,isz03b,isz03c,isz04,isz04b,isz04c}.

Quasiperiodic oscillations (QPOs) observed in the X-ray flux from accreting BHs and neutron stars provide another probe of strong-field gravity \cite{isz33,isz34}. QPOs manifest as sharp peaks in the power spectrum at frequencies ranging from a few Hz (low-frequency QPOs, LFQPOs) to hundreds of Hz (high-frequency QPOs, HFQPOs) \cite{isz35}. These oscillations are linked to the orbital motion of matter in the accretion disk, particularly near the innermost stable circular orbit (ISCO), and depend sensitively on the spacetime geometry \cite{isz36}. The epicyclic frequencies-azimuthal $\Omega_\phi$, radial $\Omega_r$, and vertical $\Omega_\theta$-and their ratios can distinguish different BH models and constrain additional parameters beyond mass and spin \cite{isz37}.

The response of BHs to external perturbations is characterized by quasinormal modes (QNMs), which are damped oscillations with complex frequencies $\omega = \omega_R + i\omega_I$ \cite{isz38,isz39}. The real part $\omega_R$ represents the oscillation frequency, while the imaginary part $\omega_I < 0$ gives the damping rate for stable modes. QNMs encode information about the BH geometry and appear in the ringdown phase of GW signals following BH mergers \cite{isz40,isz41}. For scalar field perturbations, the Klein-Gordon equation on the curved background reduces to a Schrödinger-like equation with an effective potential that depends on the metric function. The greybody factor, which quantifies the transmission probability of waves through the potential barrier, provides additional information about the spacetime structure \cite{isz42,isz43}.

The motivation for the present work arises from the need to understand how multiple matter sources simultaneously affect BH properties and observables. While individual effects of the Bardeen magnetic charge, PFDM, and CS have been studied separately \cite{isz10,isz22,isz29}, their combined influence on the same spacetime has received less attention. The charged Bardeen BH surrounded by PFDM and coupled to CS represents a configuration where regularity (from NED), dark matter effects (from PFDM), and topological defects (from CS) coexist. This multi-parameter model allows us to investigate whether these contributions can be distinguished through different observational channels.

Our aims in this work are fourfold: (i) to construct the metric function for a charged Bardeen BH with CS surrounded by PFDM and analyze its horizon structure, identifying the parameter ranges leading to non-extremal BHs, extremal BHs, single-horizon configurations, and naked singularities; (ii) to study null geodesics, compute the PS radius and shadow, and determine how $\alpha$ and $\beta$ affect these observables; (iii) to investigate the dynamics of neutral test particles, including the effective potential, specific energy, specific angular momentum, ISCO location, and QPO-related frequencies; and (iv) to examine scalar field perturbations, compute the effective potential and greybody factors, and assess how the CS and PFDM parameters modify wave propagation through the BH spacetime.

The paper is organized as follows. In Sec.~\ref{isec2}, we derive the metric function for the charged Bardeen BH with CS surrounded by PFDM, discuss the limiting cases, and analyze the horizon structure with numerical results presented in a table and figure. Section~\ref{isec3} is devoted to null geodesics, where we compute the effective potential for photons, the PS radius, and the shadow radius for various parameter combinations. In Sec.~\ref{isec4}, we study the dynamics of neutral test particles using the Hamiltonian formalism, derive the specific energy and angular momentum for circular orbits, and determine the ISCO condition. Section~\ref{isec5} examines QPOs by computing the azimuthal, radial, and vertical epicyclic frequencies, as well as the periastron precession frequency, and summarizes their parameter dependence. In Sec.~\ref{isec6}, we investigate scalar field perturbations, derive the effective potential from the Klein-Gordon equation, and compute the transmission and reflection probabilities using semi-analytical bounds. Finally, Sec.~\ref{isec7} presents our conclusions and discusses possible extensions of this work. Throughout the paper, we use geometric units with $G = c = 1$ unless otherwise stated.

\section{Spacetime Geometry and Horizon Structure of Charged Bardeen BH with PFDM and CS} \label{isec2}

In this section, we construct the spacetime geometry of a charged Bardeen BH immersed in PFDM and coupled to CS. The theoretical framework combines GR with NED, where the magnetic monopole regularizes the central singularity, while the surrounding dark matter and string cloud modify the asymptotic structure and causal properties of the spacetime.

The total action describing a charged Bardeen BH with CS surrounded by PFDM can be expressed as GR minimally coupled to NED, electromagnetic fields, dark matter, and the string cloud \cite{isz07,isz08,isz10}:
\begin{equation}
S = \int d^4x \, \sqrt{-g} \, \left(R + \mathcal{L}(F)+L_{\rm EM}\right) + S_{\rm DM} + S_{\rm CS},
\end{equation}
where $R$ is the Ricci scalar, $g = |g_{\mu\nu}|$ is the determinant of the metric tensor, $S_{\rm DM}$ represents the dark matter action, and $S_{\rm CS}$ denotes the Nambu-Goto action for the string cloud \cite{isz24}. The electromagnetic Lagrangian density is $L_{\rm EM}=-\frac{1}{4} F_{\mu\nu} F^{\mu\nu}$, and the NED Lagrangian corresponding to Bardeen's solution reads \cite{isz15,isz11}:
\begin{equation}
\mathcal{L}(F) = \frac{3}{8 \pi q^2 s} \left( \frac{\sqrt{2 q^2 F}}{2 + \sqrt{2 q^2 F}} \right)^{5/2},
\end{equation}
where $F = \frac{1}{4} F_{\mu\nu} F^{\mu\nu}$ is the electromagnetic invariant, $q$ denotes the magnetic monopole charge, $M$ is the BH mass, and $s = |q|/(2M)$ is a dimensionless parameter characterizing the magnetic charge strength.

Variation of the action yields the Einstein field equations:
\begin{equation}
    G_{\mu\nu}=8 \pi \left(T^{\rm NED}_{\,\,\mu\nu}+T^{\rm EM}_{\mu\nu}+T^{\rm DM}_{\,\,\mu\nu}\right)+T^{\rm CS}_{\,\,\mu\nu}.\label{EFE}
\end{equation}

The energy-momentum tensor of PFDM takes the diagonal form \cite{isz18,isz20}:
\begin{equation}
T^{\rm DM}_{\,\,\mu\nu} = \mathrm{diag}\left(-\mathcal{E}_{\rm DM},\, P_{r\,\rm DM},\, P_{\theta\,\rm DM},\, P_{\phi\,\rm DM}\right),
\label{tensor}
\end{equation}
with the energy density and pressure components given by
\begin{equation}
\mathcal{E}_{\rm DM} = -P_{r\,\rm DM}= -\frac{\beta}{8\pi r^3},
\label{energy_density}
\end{equation}
and
\begin{equation}
P_{\theta\,\rm DM}= -\frac{\beta}{16\pi r^3}=P_{\phi\,\rm DM}.
\label{pressures}
\end{equation}
Here, $\beta > 0$ is the PFDM parameter that quantifies the dark matter intensity surrounding the BH. The anisotropic pressure profile, with $P_r \neq P_\theta = P_\phi$, distinguishes PFDM from isotropic perfect fluid models and leads to logarithmic corrections in the metric function \cite{isz19,isz21}.

The CS contribution arises from the Nambu-Goto action \cite{isz24,isz28}:
\begin{equation}
    S_{\rm CS}=\int \sqrt{-\gamma}\,\mathcal{M}\,d\lambda^0\,d\lambda^1=\int \mathcal{M}\sqrt{-\frac{1}{2}\,\Sigma^{\mu \nu}\,\Sigma_{\mu\nu}}\,d\lambda^0\,d\lambda^1,\label{act1}
\end{equation}
where $\mathcal{M}$ is a dimensionless constant characterizing the string tension, and $(\lambda^0, \lambda^1)$ are the timelike and spacelike worldsheet coordinates, respectively \cite{JLS1960}. The induced metric determinant on the string worldsheet is $\gamma = g^{\mu\nu}\frac{\partial x^\mu}{\partial \lambda^a}\frac{\partial x^\nu}{\partial \lambda^b}$, and the bivector $\Sigma_{\mu\nu}=\epsilon^{ab}\frac{\partial x^\mu}{\partial \lambda^a}\frac{\partial x^\nu}{\partial \lambda^b}$ encodes the worldsheet orientation, with $\epsilon^{ab}$ being the two-dimensional Levi-Civita symbol satisfying $\epsilon^{01} = -\epsilon^{10} = 1$.

The resulting energy-momentum tensor for the CS reads:
\begin{equation}
   T_{\mu\nu}^{\rm CS}=2 \frac{\partial}{\partial g_{\mu \nu}}\mathcal{M}\sqrt{-\frac{1}{2}\Sigma^{\mu \nu}\,\Sigma_{\mu\nu}} =\frac{\rho^{\rm CS} \,\Sigma_{\alpha\nu}\, \,\Sigma_{\mu}^\alpha }{\sqrt{-\gamma}}, \label{act2}
\end{equation}
where $\rho^{\rm CS}$ is the proper string cloud density. The non-vanishing components are \cite{isz27}:
\begin{equation}
    T^{t\,(\rm CS)}_{t}=\rho^{\rm CS}=\frac{\alpha}{r^2}=T^{r\,(\rm CS)}_{r},\quad T^{\theta\,(\rm CS)}_{\theta}=T^{\phi\,(\rm CS)}_{\phi}= 0,\label{act3}
\end{equation}
where $\alpha \in [0,1)$ is the CS parameter representing the integrated effect of the string network on the spacetime geometry.

The line element for a static, spherically symmetric charged Bardeen BH with CS surrounded by PFDM is:
\begin{equation}
ds^2 = -f(r)\,dt^2 + \frac{dr^2}{f(r)} + r^2\left(d\theta^2 + \sin^2\theta\,d\phi^2\right),\label{metric}
\end{equation}
where the metric function combines contributions from all matter sources \cite{isz18,isz24,isz29}:
\begin{equation}
f(r) = 1 - \alpha - \frac{2Mr^2}{\left(q^2 + r^2\right)^{3/2}} + \frac{Q^2}{r^2} + \frac{\beta}{r}\ln\!\frac{r}{|\beta|}.\label{function}
\end{equation}
The five parameters governing the spacetime geometry are: $M$ (mass), $q$ (magnetic monopole charge from NED), $Q$ (electric charge), $\alpha$ (CS parameter), and $\beta$ (PFDM parameter).

\subsection{Limiting Cases}

The metric function in Eq.~\eqref{function} reduces to several well-known BH solutions under appropriate parameter choices:

\begin{itemize}
    \item \textbf{Bardeen BH with PFDM} ($\alpha = 0$, $Q = 0$): The metric function becomes
    \begin{equation}
        f(r) = 1 - \frac{2Mr^2}{\left(q^2 + r^2\right)^{3/2}} + \frac{\beta}{r}\ln\!\frac{r}{|\beta|},
        \label{function1}
    \end{equation}
    describing a regular Bardeen BH immersed in PFDM \cite{isz18}.

    \item \textbf{RN BH with CS and PFDM} ($q = 0$): The metric function takes the form
    \begin{equation}
        f(r) = 1 - \alpha - \frac{2M}{r} + \frac{Q^2}{r^2} + \frac{\beta}{r}\ln\!\frac{r}{|\beta|},
        \label{function2}
    \end{equation}
    representing an RN BH with CS surrounded by PFDM \cite{DVS2025}.

    \item \textbf{Schwarzschild BH with CS and PFDM} ($q = 0$, $Q = 0$): The metric function simplifies to
    \begin{equation}
        f(r) = 1 - \alpha - \frac{2M}{r} + \frac{\beta}{r}\ln\!\frac{r}{|\beta|},
        \label{function3}
    \end{equation}
    corresponding to a Schwarzschild BH with CS in PFDM \cite{isz30}.

    \item \textbf{Bardeen BH with CS} ($Q = 0$, $\beta = 0$): The metric function reduces to
    \begin{equation}
        f(r) = 1 - \alpha - \frac{2Mr^2}{\left(q^2 + r^2\right)^{3/2}},\label{function4}
    \end{equation}
    describing the Bardeen BH coupled with a CS \cite{isz29,isz14}.
\end{itemize}

\subsection{Horizon Structure and Causal Classification} \label{sec2.5}

The event horizons are located at the positive real roots of the equation $f(r_h) = 0$. Due to the presence of the logarithmic PFDM term and the Bardeen-type magnetic charge, the horizon equation
\begin{equation}
1 - \alpha - \frac{2Mr_h^2}{\left(q^2 + r_h^2\right)^{3/2}} + \frac{Q^2}{r_h^2} + \frac{\beta}{r_h}\ln\!\frac{r_h}{|\beta|} = 0
\label{horizon_eq}
\end{equation}
does not admit closed-form analytical solutions and must be solved numerically. The number and nature of horizons depend critically on the parameter space $(M, q, Q, \alpha, \beta)$.

We classify the spacetime configurations based on the number and nature of positive real roots of $f(r_h) = 0$ \cite{sec2is08}. When two distinct positive roots $r_- < r_+$ exist, corresponding to the Cauchy (inner) and event (outer) horizons respectively, the configuration is termed a non-extremal BH with positive surface gravity $\kappa = f'(r_+)/2$. The extremal BH configuration occurs when these two horizons coincide at a degenerate radius $r_- = r_+ = r_{\rm ext}$, satisfying both $f(r_{\rm ext}) = 0$ and $f'(r_{\rm ext}) = 0$ simultaneously, which results in vanishing Hawking temperature. A single-horizon BH arises when only one positive root exists, analogous to the Schwarzschild case, or when the magnetic charge and PFDM effects combine to eliminate the inner horizon. Finally, the absence of positive real roots leads to a naked singularity where the central region-a regular core for Bardeen-type solutions-becomes globally visible, potentially violating cosmic censorship \cite{sec2is09}.

Table~\ref{tab:horizons} presents the horizon radii for representative parameter combinations, illustrating the transition between different causal structures. The table includes all four horizon configurations: single-horizon BHs (Schwarzschild-type), non-extremal BHs with distinct inner and outer horizons, extremal BHs where the two horizons coincide at a degenerate radius $r_{\rm ext}$ satisfying both $f(r_{\rm ext}) = 0$ and $f'(r_{\rm ext}) = 0$, and naked singularities where no horizon exists. The metric function behavior for selected configurations is displayed in Fig.~\ref{fig:metric_function}.

\setlength{\tabcolsep}{10pt}
\renewcommand{\arraystretch}{1.5}
\begin{longtable}{|c|c|c|c|c|c|}
\hline
\rowcolor{orange!50}
\textbf{$\alpha$} & \textbf{$\beta/M$} & \textbf{$q/M$} & \textbf{$Q/M$} & \textbf{Horizon(s) $[r_h/M]$} & \textbf{Configuration} \\
\hline
\endfirsthead
\hline
\rowcolor{orange!50}
\textbf{$\alpha$} & \textbf{$\beta/M$} & \textbf{$q/M$} & \textbf{$Q/M$} & \textbf{Horizon(s) $[r_h/M]$} & \textbf{Configuration} \\
\hline
\endhead
\hline
\endfoot
\hline
\endlastfoot
0.00 & 0.0 & 0.0 & 0.00 & $[2.0000]$ & Single-horizon BH \\
\hline
0.10 & 0.0 & 0.0 & 0.00 & $[2.2222]$ & Single-horizon BH \\
\hline
0.00 & 0.0 & 0.5 & 0.00 & $[0.3262,\ 1.7860]$ & Non-extremal BH \\
\hline
0.10 & 0.0 & 0.5 & 0.00 & $[0.2978,\ 2.0352]$ & Non-extremal BH \\
\hline
0.00 & 1.0 & 0.1 & 1.00 & $[0.4559,\ 0.9679]$ & Non-extremal BH \\
\hline
0.10 & 1.0 & 0.1 & 1.00 & $[0.4374,\ 1.0796]$ & Non-extremal BH \\
\hline
0.00 & 0.651 & 0.2 & 0.85 & $[0.7252]$ & Extremal BH \\
\hline
0.10 & 0.604 & 0.15 & 0.90 & $[0.7566]$ & Extremal BH \\
\hline
0.15 & 1.2 & 0.1 & 1.00 & $[0.3314,\ 1.3105]$ & Non-extremal BH \\
\hline
0.20 & 1.5 & 0.1 & 1.00 & $[0.2428,\ 1.5899]$ & Non-extremal BH \\
\hline
0.00 & 0.5 & 0.1 & 0.50 & $[0.1547,\ 1.3102]$ & Non-extremal BH \\
\hline
0.10 & 0.8 & 0.2 & 0.70 & $[0.2940,\ 1.2962]$ & Non-extremal BH \\
\hline
0.25 & 1.8 & 0.1 & 1.00 & $[0.1888,\ 1.8598]$ & Non-extremal BH \\
\hline
0.30 & 2.0 & 0.5 & 0.50 & $[0.0298,\ 2.2071]$ & Non-extremal BH \\
\hline
0.00 & 0.5 & 0.8 & 1.50 & $[\,]$ & Naked singularity \\
\hline
0.00 & 0.8 & 0.2 & 0.95 & $[\,]$ & Naked singularity \\
\hline
0.05 & 1.4 & 0.1 & 1.00 & $[0.2707,\ 1.3580]$ & Non-extremal BH \\
\\
\caption{Horizon radii $r_h/M$ for charged Bardeen BH with CS and PFDM for various parameter combinations. The Schwarzschild limit ($\alpha = \beta = q = Q = 0$) yields $r_h = 2M$, while the CS parameter $\alpha > 0$ shifts the outer horizon outward. Non-extremal configurations exhibit two distinct horizons $r_- < r_+$, whereas extremal BHs possess a single degenerate horizon where $f(r_{\rm ext}) = f'(r_{\rm ext}) = 0$ and the Hawking temperature vanishes. Naked singularities form when the charge parameters $(q, Q)$ become sufficiently large relative to the mass, preventing horizon formation.}
\label{tab:horizons}
\end{longtable}

\begin{figure}[ht!]
    \centering
    \includegraphics[width=0.85\linewidth]{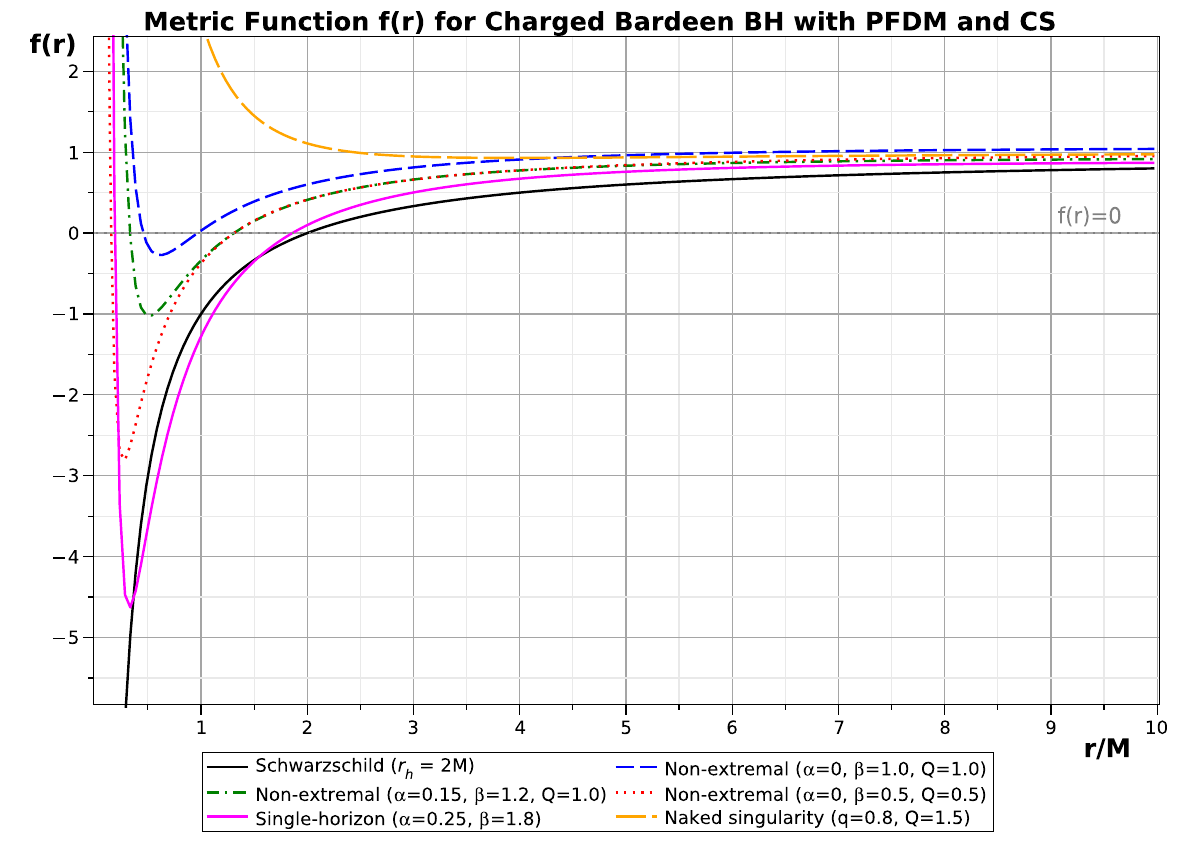}
    \caption{Metric function $f(r)$ versus $r/M$ for the charged Bardeen BH with PFDM and CS. Six representative configurations are shown: Schwarzschild reference with $r_h = 2M$ (black solid); non-extremal BH with $(\alpha, \beta/M, q/M, Q/M) = (0, 1.0, 0.1, 1.0)$ having horizons at $r_h/M = [0.4559, 0.9679]$ (blue dashed); non-extremal BH with $(0.15, 1.2, 0.1, 1.0)$ and $r_h/M = [0.3314, 1.3105]$ (green dash-dotted); non-extremal BH with $(0, 0.5, 0.1, 0.5)$ and $r_h/M = [0.1547, 1.3102]$ (red dotted); single-horizon configuration with $(0.25, 1.8, 0.1, 1.0)$ showing $r_h/M = [0.1888, 1.8598]$ (magenta solid); and naked singularity with $(0, 0.5, 0.8, 1.5)$ where $f(r) > 0$ for all $r > 0$ (orange dashed). The horizontal gray line marks $f(r) = 0$, and intersections indicate horizon locations.}
    \label{fig:metric_function}
\end{figure}

The numerical results in Table~\ref{tab:horizons} reveal several important features of the horizon structure:

\textbf{(i) Schwarzschild and Schwarzschild-CS limits:} In the absence of magnetic charge, electric charge, and PFDM ($q = Q = \beta = 0$), the metric reduces to Schwarzschild form with a single horizon at $r_h = 2M/(1-\alpha)$. For $\alpha = 0$, we recover $r_h = 2M$, while $\alpha = 0.1$ yields $r_h \approx 2.222M$, confirming that the CS parameter shifts the horizon outward.

\textbf{(ii) Bardeen effect:} The magnetic monopole charge $q$ introduces an inner (Cauchy) horizon, transforming the single-horizon Schwarzschild geometry into a non-extremal two-horizon structure. For instance, with $q/M = 0.5$ and $\alpha = \beta = Q = 0$, we find $r_-/M = 0.3262$ and $r_+/M = 1.7860$.

\textbf{(iii) PFDM influence:} The logarithmic term from PFDM modifies both horizon locations. Increasing $\beta/M$ generally widens the gap between inner and outer horizons. Comparing rows 5--8 in Table~\ref{tab:horizons}, we observe that as $\beta/M$ increases from 1.0 to 1.5 (with corresponding $\alpha$ increase), the outer horizon expands from $r_+/M = 0.9679$ to $r_+/M = 1.5899$.

\textbf{(iv) Naked singularity formation:} When the electric charge $Q$ and magnetic charge $q$ become sufficiently large relative to the mass, the repulsive electromagnetic contributions dominate, preventing horizon formation. The configurations with $(q/M, Q/M) = (0.8, 1.5)$ and $(0.2, 0.95)$ yield naked singularities, as confirmed by the absence of roots in Table~\ref{tab:horizons} and the positive-definite behavior of the orange curve in Fig.~\ref{fig:metric_function}.

The asymptotic behavior of the metric function at large distances is:
\begin{equation}
\lim_{r \to \infty} f(r) = 1 - \alpha \neq 1,
\label{asymptotic}
\end{equation}
indicating that the spacetime is asymptotically non-flat due to the CS contribution. This deviation from asymptotic flatness has important implications for the PS radius and shadow calculations presented in subsequent sections \cite{isz31}.

\section{Observable Signatures: Photon Sphere and BH Shadow} \label{isec3}

In this section, we investigate null geodesics for the selected charged Bardeen BH with PFDM and CS, focusing on the effective potential, radial force experienced by photons, PS radius, and BH shadow. We explore and analyze how the geometric parameters $(\alpha, \beta, q, Q)$ that alters the space-time curvature influence the photon dynamics and observable signatures.

The null geodesics are studied through the Lagrangian formalism, where the Lagrangian density in terms of the metric tensor $g_{\mu\nu}$ is given by \cite{sec4is02, sec2is08}:
\begin{equation}
    \mathcal{L}=\frac{1}{2} g_{\mu\nu} \dot{x}^{\mu} \dot{x}^{\nu},\label{bb1}
\end{equation}
where the overdot denotes differentiation with respect to an affine parameter $\lambda$. For the spacetime metric in Eq.~\eqref{metric}, the metric tensor components are
\begin{equation}
    g_{\mu\nu}=\mathrm{diag}\left(-f(r),\, \frac{1}{f(r)},\, r^2,\, r^2 \sin^2\theta\right)\qquad (\mu,\nu=0,\ldots,3).\label{bb2}
\end{equation}

The Lagrangian density simplifies to
\begin{equation}
    \mathcal{L}=\frac{1}{2}\left[-f(r)\,\dot{t}^2+\frac{\dot{r}^2}{f(r)}+r^2\,\dot{\theta}^2+ r^2 \sin^2\theta\,\dot{\phi}^2\right].\label{bb3}
\end{equation}
Since the Lagrangian is independent of the coordinates $t$ and $\phi$, two conserved quantities emerge from the Euler-Lagrange equations:
\begin{equation}
    \mathrm{E}=-\frac{\partial \mathcal{L}}{\partial \dot{t}}=f(r)\,\dot{t}.\label{bb4}
\end{equation}
And
\begin{equation}
    \mathrm{L}=-\frac{\partial \mathcal{L}}{\partial \dot{\phi}}=r^2 \sin^2\theta\,\dot{\phi},\label{bb5}
\end{equation}
where $\mathrm{E}$ and $\mathrm{L}$ represent the conserved energy and angular momentum of the photon particles, respectively.

Restricting to the equatorial plane ($\theta=\pi/2$, $\dot{\theta}=0$), the radial equation of motion becomes
\begin{equation}
    \dot{r}^2+\frac{\mathrm{L}^2}{r^2}\,f(r)=\mathrm{E}^2.\label{bb6}
\end{equation}
This can be recast as a one-dimensional energy equation $\dot{r}^2+V_{\rm eff}=\mathrm{E}^2$, where the effective potential is
\begin{equation}
    V_{\rm eff}=\frac{\mathrm{L}^2}{r^2}\,f(r)=\frac{\mathrm{L}^2}{r^2} \left[1 - \alpha - \frac{2Mr^2}{\left(q^2 + r^2\right)^{3/2}} + \frac{Q^2}{r^2} + \frac{\beta}{r}\ln\frac{r}{\beta}\right].\label{bb7}
\end{equation}

The effective potential in Eq.~\eqref{bb7} depends explicitly on all five spacetime parameters: the CS parameter $\alpha$, the PFDM parameter $\beta$, the magnetic monopole charge $q$, the electric charge $Q$, and the BH mass $M$. Each parameter modifies the spacetime geometry and reshapes the potential barrier experienced by photons, affecting the PS location, shadow radius, and light bending characteristics.

\begin{figure}[ht!]
    \centering
    \includegraphics[width=0.45\linewidth]{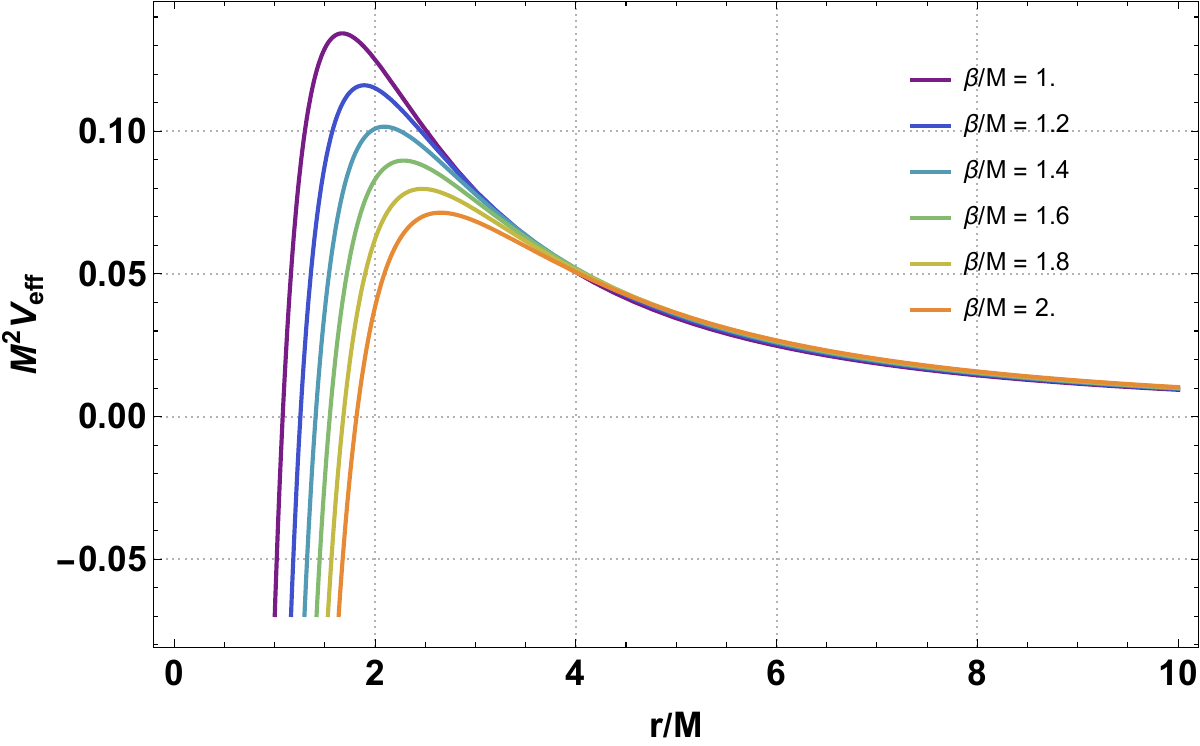}\quad
    \includegraphics[width=0.45\linewidth]{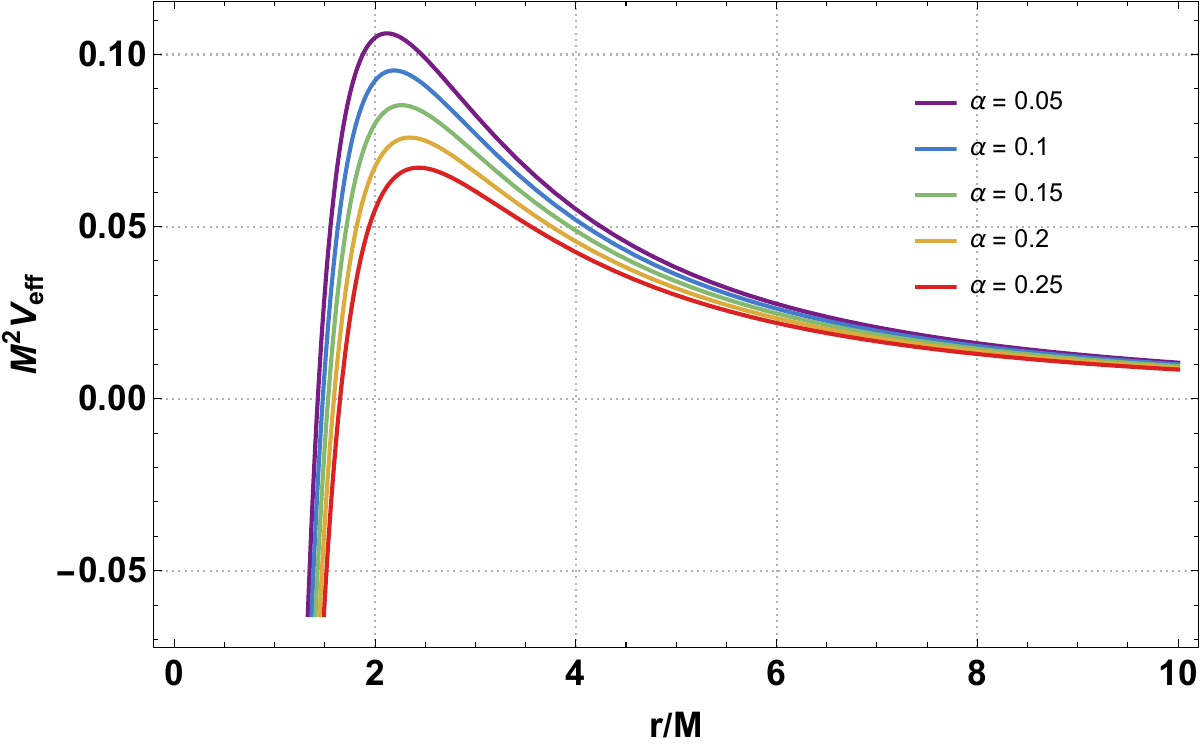}\\
    (a) $\alpha=0.1$ \hspace{6cm} (b) $\beta/M=1.5$
    \caption{Effective potential $M^2 V_{\rm eff}$ governing photon dynamics as a function of $r/M$ for various values of $\alpha$ and $\beta/M$. Panel (a): varying $\beta/M = 1.0, 1.2, 1.4, 1.6, 1.8, 2.0$ with fixed $\alpha = 0.1$. Panel (b): varying $\alpha = 0.05, 0.1, 0.15, 0.2, 0.25$ with fixed $\beta/M = 1.5$. Other parameters: $Q/M=1$, $q/M=0.1$. Increasing both $\beta/M$ and $\alpha$ suppresses the potential peak, reducing the gravitational barrier for photons.}
    \label{fig:null}
\end{figure}

Figure~\ref{fig:null} displays the effective potential as a function of $r/M$ for various parameter combinations. In panel (a), increasing $\beta/M$ progressively lowers the potential peak, indicating that PFDM reduces the gravitational barrier. Panel (b) shows that larger values of $\alpha$ similarly suppress the peak. This behavior implies that both the CS and PFDM weaken the photon trapping capability of the BH, allowing photons with lower impact parameters to escape.

The effective radial force experienced by photons is defined as the negative gradient of the effective potential:
\begin{equation}
    F_{\rm eff}=-\frac{1}{2}\frac{\partial V_{\rm eff}}{\partial r}.\label{bb8}
\end{equation}
Substituting Eq.~\eqref{bb7} yields
\begin{equation}
    F_{\rm eff}=\frac{\mathrm{L}^2}{r^3} \left[1-\alpha-\frac{3Mr^4}{(q^2+r^2)^{5/2}}+\frac{2Q^2}{r^2}+\frac{\beta}{2r}\left(3\ln\frac{r}{\beta}-1\right)\right].\label{bb9}
\end{equation}

\begin{figure}[ht!]
    \centering
    \includegraphics[width=0.45\linewidth]{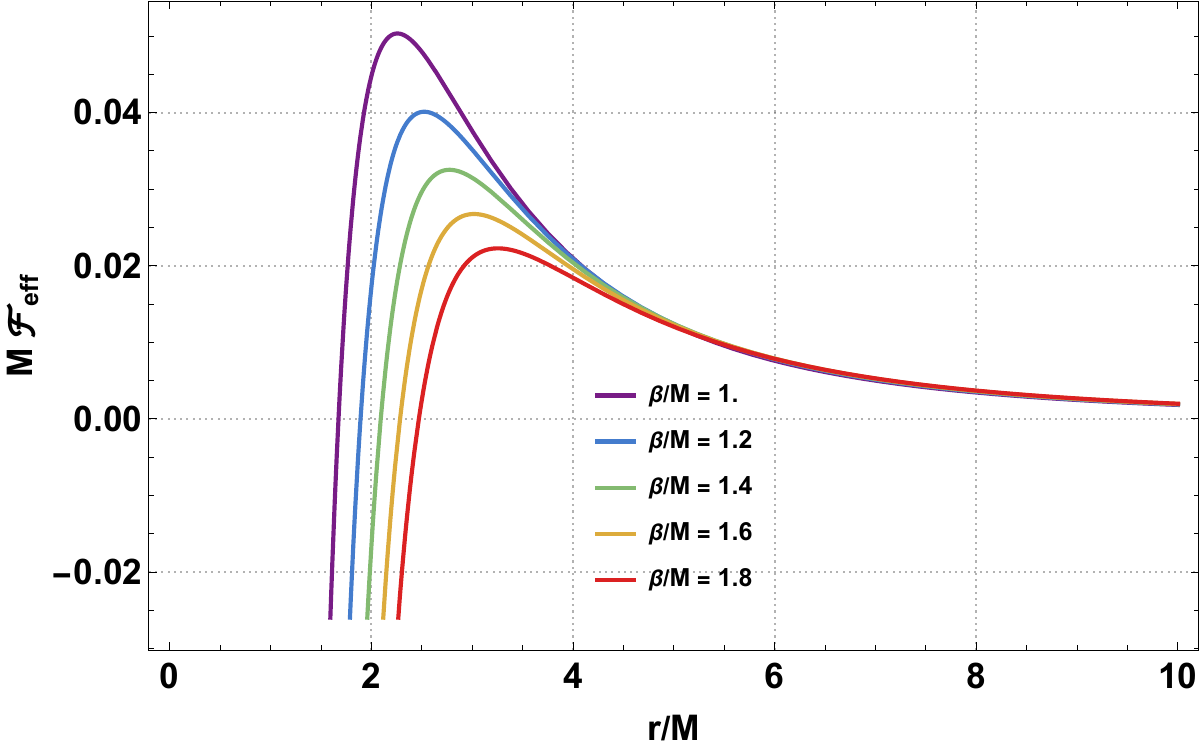}\quad
    \includegraphics[width=0.45\linewidth]{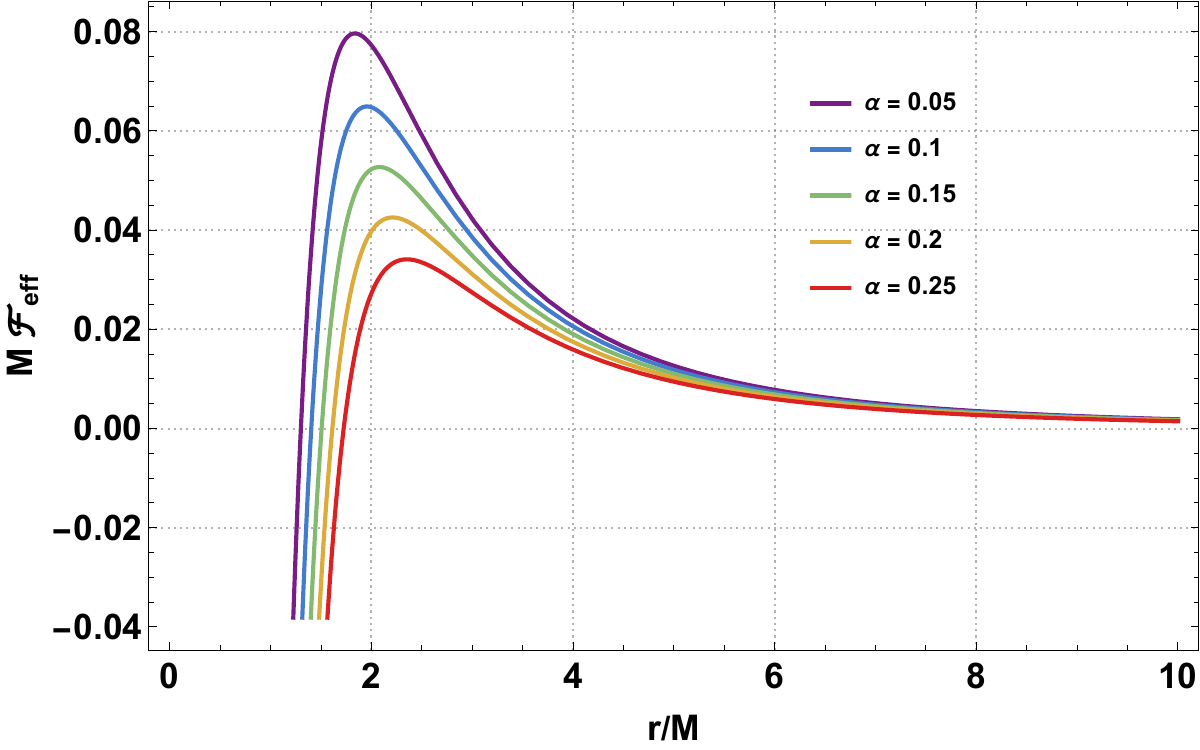}\\
    (a) $\alpha=0.1$ \hspace{6cm} (b) $\beta/M=0.8$
    \caption{Effective radial force $M\,F_{\rm eff}$ on photons as a function of $r/M$. Panel (a): varying $\beta/M = 1.0, 1.2, 1.4, 1.6, 1.8$ with fixed $\alpha = 0.1$. Panel (b): varying $\alpha = 0.05, 0.1, 0.15, 0.2, 0.25$ with fixed $\beta/M = 0.8$. Other parameters: $Q/M=1$, $q/M=0.1$, $\mathrm{L}/M=1$. The force vanishes at the PS radius and approaches zero as $r \to \infty$.}
    \label{fig:force-photons}
\end{figure}

Figure~\ref{fig:force-photons} illustrates the radial force behavior. The force changes sign at the PS radius, where $F_{\rm eff} = 0$ corresponds to the unstable circular orbit condition. Increasing $\beta/M$ and $\alpha$ reduces the force magnitude, consistent with the weakened potential barrier observed in Fig.~\ref{fig:null}. At large distances ($r \to \infty$), the radial force vanishes as expected for asymptotically approached spacetimes.

Circular photon orbits satisfy the conditions $\dot{r}=0$ and $\ddot{r}=0$, which lead to
\begin{equation}
    \mathrm{E}^2=\frac{\mathrm{L}^2}{r^2}\,f(r),\label{bb10}
\end{equation}
\begin{equation}
    \frac{\partial V_{\rm eff}}{\partial r}=0.\label{bb11}
\end{equation}
Substituting Eq.~\eqref{bb7} into Eq.~\eqref{bb11} gives the PS condition:
\begin{equation}
    1-\alpha-\frac{3Mr^4}{(q^2+r^2)^{5/2}}+\frac{2Q^2}{r^2}+\frac{\beta}{2r}\left(3\ln\frac{r}{\beta}-1\right)=0.\label{bb12}
\end{equation}
Due to the logarithmic term, Eq.~\eqref{bb12} does not admit a closed-form solution and must be solved numerically for the PS radius $r_{\rm ph}$.

Tables~\ref{tab:1} and \ref{tab:2} present numerical values of the PS radius $r_{\rm ph}/M$ for various combinations of $\alpha$ and $\beta/M$, with fixed charge parameters $Q/M=1$ and two values of the magnetic monopole charge: $q/M=0.1$ (Table~\ref{tab:1}) and $q/M=0.2$ (Table~\ref{tab:2}).


\setlength{\tabcolsep}{12pt}
\renewcommand{\arraystretch}{1.6}
\begin{table}[ht!]
\centering
\begin{tabular}{|>{\columncolor{orange!50}}c|c|c|c|c|c|c|c|}
\hline
\rowcolor{orange!50}
\diagbox[innerwidth=1.5cm, height=1.6cm, linecolor=black, font=\normalsize]{\hspace{0.9em}\raisebox{0.1em}{$\alpha$}}{\raisebox{-0.1em}{$\beta/M$}\hspace{0.0em}} & \textbf{0.6} & \textbf{0.8} & \textbf{1.0} & \textbf{1.2} & \textbf{1.4} & \textbf{1.6} & \textbf{1.8} \\
\hline
\textbf{0.05} & 1.16212 & 0.77510 & 1.59616 & 1.81988 & 2.02181 & 2.21313 & 2.39822 \\
\hline
\textbf{0.10} & 1.16212 & 0.74540 & 1.67327 & 1.89177 & 2.09245 & 2.28413 & 2.47040 \\
\hline
\textbf{0.15} & 1.16212 & 0.72192 & 1.75571 & 1.96898 & 2.16829 & 2.36023 & 2.54765 \\
\hline
\textbf{0.20} & 1.16212 & 0.70238 & 1.84435 & 2.05227 & 2.24999 & 2.44207 & 2.63059 \\
\hline
\textbf{0.25} & 1.45975 & 1.72463 & 1.94021 & 2.14249 & 2.33836 & 2.53042 & 2.71994 \\
\hline
\end{tabular}
\caption{PS radius $r_{\rm ph}/M$ for various $\alpha$ and $\beta/M$ with $Q/M=1$, $q/M=0.1$.}
\label{tab:1}
\end{table}

\setlength{\tabcolsep}{12pt}
\renewcommand{\arraystretch}{1.6}
\begin{table}[ht!]
\centering
\begin{tabular}{|>{\columncolor{orange!50}}c|c|c|c|c|c|c|c|}
\hline
\rowcolor{orange!50}
\diagbox[innerwidth=1.5cm, height=1.6cm, linecolor=black, font=\normalsize]{\hspace{0.9em}\raisebox{0.1em}{$\alpha$}}{\raisebox{-0.1em}{$\beta/M$}\hspace{0.0em}} & \textbf{0.6} & \textbf{0.8} & \textbf{1.0} & \textbf{1.2} & \textbf{1.4} & \textbf{1.6} & \textbf{1.8} \\
\hline
\textbf{0.05} & 1.33031 & 1.12958 & 0.70955 & 1.76617 & 1.98407 & 2.18444 & 2.37538 \\
\hline
\textbf{0.10} & 1.33031 & 1.12958 & 1.59009 & 1.84070 & 2.05601 & 2.25621 & 2.44807 \\
\hline
\textbf{0.15} & 1.33031 & 1.31015 & 0.68098 & 1.92040 & 2.13311 & 2.33307 & 2.52583 \\
\hline
\textbf{0.20} & 1.33031 & 0.92523 & 1.77433 & 2.00603 & 2.21604 & 2.41567 & 2.60927 \\
\hline
\textbf{0.25} & 1.33031 & 0.88167 & 1.87557 & 2.09849 & 2.30562 & 2.50476 & 2.69913 \\
\hline
\end{tabular}
\caption{PS radius $r_{\rm ph}/M$ for various $\alpha$ and $\beta/M$ with $Q/M=1$, $q/M=0.2$.}
\label{tab:2}
\end{table}

Several features emerge from the numerical data. For $\beta/M \geq 1.2$, the PS radius increases monotonically with both $\alpha$ and $\beta/M$, indicating that the CS and PFDM push the unstable photon orbit outward. For instance, at $(\alpha, \beta/M) = (0.05, 1.2)$ with $q/M = 0.1$, we obtain $r_{\rm ph}/M = 1.81988$, which increases to $r_{\rm ph}/M = 2.71994$ at $(\alpha, \beta/M) = (0.25, 1.8)$—a relative increase of approximately $49\%$. The non-monotonic behavior observed at smaller $\beta/M$ values (e.g., $\beta/M = 0.6, 0.8, 1.0$) reflects the complex interplay between the PFDM logarithmic term, the Bardeen magnetic charge, and the electric charge contributions in Eq.~\eqref{bb12}. Comparing Tables~\ref{tab:1} and \ref{tab:2}, we note that increasing the magnetic monopole charge from $q/M = 0.1$ to $q/M = 0.2$ generally raises the PS radius at small $\beta/M$, consistent with the regularizing effect of the Bardeen-type charge on the spacetime geometry. Since the PS radius directly determines the BH shadow size through Eq.~\eqref{bb16}, these parameter-dependent variations in $r_{\rm ph}$ translate into potentially distinguishable shadow signatures for different CS and PFDM configurations.

\begin{figure}[ht!]
    \centering
    \includegraphics[width=0.55\linewidth]{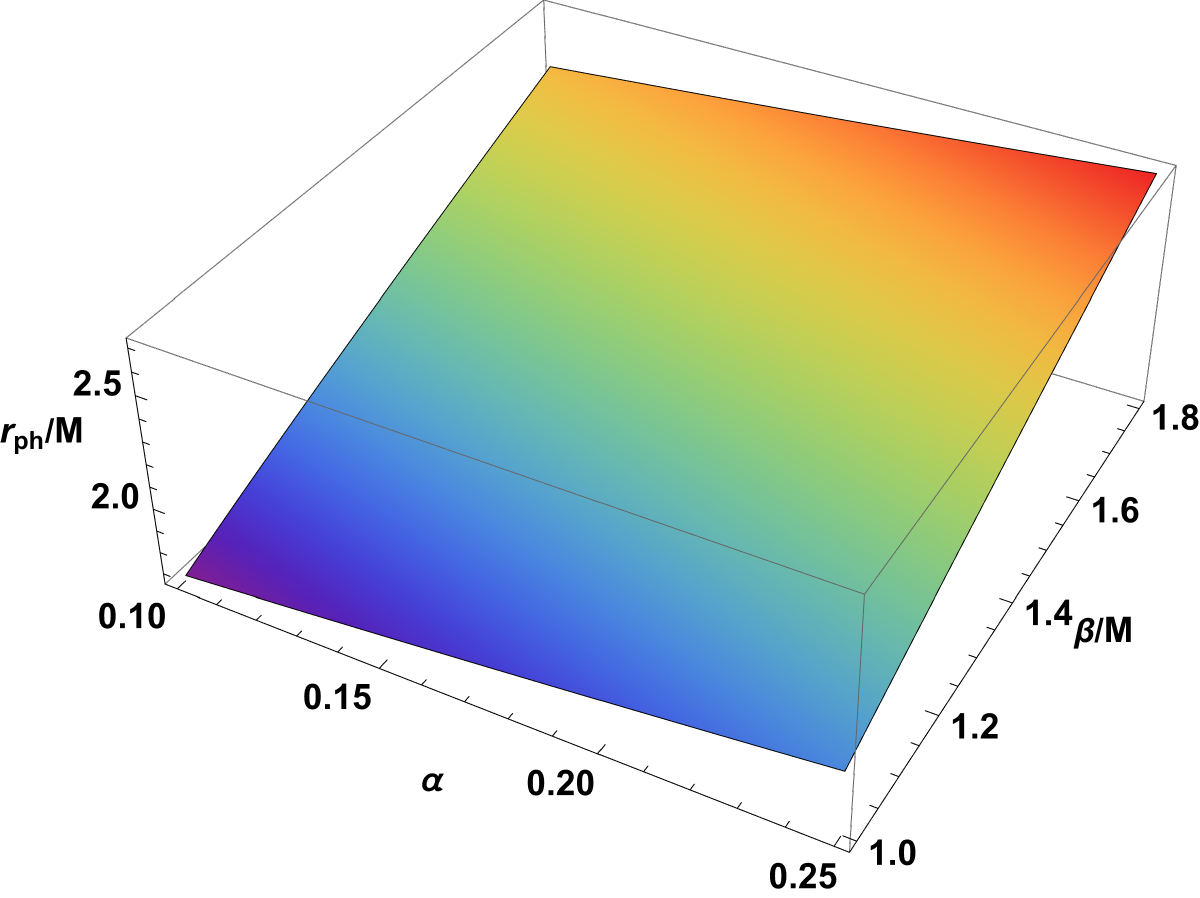}
    \caption{Three-dimensional surface plot of the PS radius $r_{\rm ph}/M$ as a function of $\alpha \in [0.05, 0.25]$ and $\beta/M \in [1.0, 1.8]$ with fixed $Q/M=1$, $q/M=0.1$. The PS radius increases monotonically with both parameters.}
    \label{fig:1}
\end{figure}

Figure~\ref{fig:1} presents a 3D visualization of the PS radius dependence on $\alpha$ and $\beta/M$. The surface shows a monotonic increase in $r_{\rm ph}/M$ with both parameters, confirming that the combined effects of CS and PFDM expand the photon capture region.

For asymptotically non-flat spacetimes where $\lim_{r \to \infty} f(r) = 1 - \alpha \neq 1$, the shadow radius for an observer at position $r_O$ is given by \cite{isz31}
\begin{equation}
    R_{\rm sh}=r_{\rm ph}\sqrt{\frac{f(r_O)}{f(r_{\rm ph})}}.\label{bb14}
\end{equation}
Substituting the metric function yields
\begin{equation}
    R_{\rm sh}=r_{\rm ph}\sqrt{\frac{1 - \alpha - \frac{2Mr^2_O}{(q^2 + r^2_O)^{3/2}} + \frac{Q^2}{r^2_O} + \frac{\beta}{r_O}\ln\frac{r_O}{\beta}}{1 - \alpha - \frac{2Mr^2_{\rm ph}}{(q^2 + r^2_{\rm ph})^{3/2}} + \frac{Q^2}{r^2_{\rm ph}} + \frac{\beta}{r_{\rm ph}}\ln\frac{r_{\rm ph}}{\beta}}}.\label{bb15}
\end{equation}
For a distant observer ($r_O \to \infty$), the shadow radius simplifies to
\begin{equation}
    R_{\rm sh}=(1-\alpha)^{1/2}\,\frac{r_{\rm ph}}{\sqrt{f(r_{\rm ph})}},\label{bb16}
\end{equation}
where $r_{\rm ph}$ is determined from Eq.~\eqref{bb12}.

\setlength{\tabcolsep}{12pt}
\renewcommand{\arraystretch}{1.6}
\begin{table}[ht!]
\centering
\begin{tabular}{|>{\columncolor{orange!50}}c|c|c|c|c|c|c|c|}
\hline
\rowcolor{orange!50}
\diagbox[innerwidth=1.5cm, height=1.6cm, linecolor=black, font=\normalsize]{\hspace{0.9em}\raisebox{0.1em}{$\alpha$}}{\raisebox{-0.1em}{$\beta/M$}\hspace{0.0em}} & \textbf{0.6} & \textbf{0.8} & \textbf{1.0} & \textbf{1.2} & \textbf{1.4} & \textbf{1.6} & \textbf{1.8} \\
\hline
\textbf{0.05} & 1.97263 & 2.28304 & 2.49184 & 2.69713 & 2.89443 & 3.08863 & 3.28136 \\
\hline
\textbf{0.10} & 2.08458 & 2.38067 & 2.58903 & 2.78466 & 2.97669 & 3.16773 & 3.36386 \\
\hline
\textbf{0.15} & 2.23548 & 2.50615 & 2.69263 & 2.87782 & 3.06388 & 3.25124 & 3.43973 \\
\hline
\textbf{0.20} & 2.45194 & 2.67418 & 2.80349 & 2.97723 & 3.15649 & 3.33952 & 3.52516 \\
\hline
\textbf{0.25} & 2.66973 & 2.91269 & 2.92261 & 3.08361 & 3.25506 & 3.43297 & 3.61516 \\
\hline
\end{tabular}
\caption{Shadow radius $R_{\rm sh}/M$ for various $\alpha$ and $\beta/M$ with $Q/M=1$, $q/M=0.1$.}
\label{tab:shadow2}
\end{table}

\begin{figure}[ht!]
    \centering
    \includegraphics[width=0.55\linewidth]{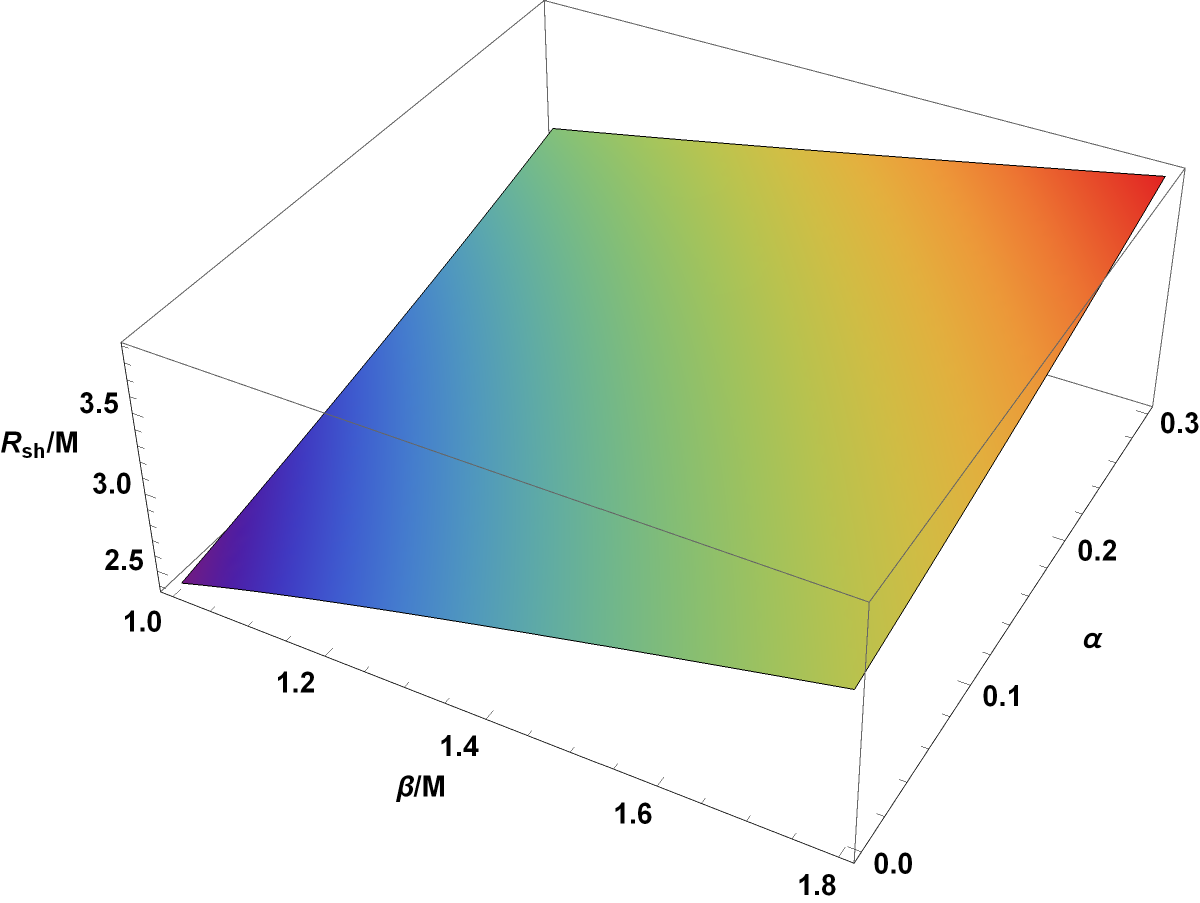}
    \caption{Three-dimensional surface plot of the shadow radius $R_{\rm sh}/M$ as a function of $\alpha \in [0.05, 0.3]$ and $\beta/M \in [1.0, 1.8]$ with fixed $Q/M=1$, $q/M=0.3$. The shadow radius increases monotonically with both $\alpha$ and $\beta/M$.}
    \label{fig:shadow2}
\end{figure}

Table~\ref{tab:shadow2} and Fig.~\ref{fig:shadow2} present the shadow radius dependence on the CS and PFDM parameters. The shadow radius $R_{\rm sh}/M$ increases monotonically with both $\alpha$ and $\beta/M$. Physically, the CS enhances the effective gravitational attraction by reducing the asymptotic metric value, while PFDM introduces logarithmic corrections that modify the spacetime at large distances. Both effects push the PS outward and enlarge the shadow, making these parameters potentially distinguishable through EHT observations of supermassive BH shadows \cite{isz03,isz04}.

The angular velocity of photons in circular orbits is defined by
\begin{equation}
    \Omega^{\rm null}_{\phi}(r)=\frac{\dot{\phi}}{\dot{t}}=\frac{\sqrt{f(r)}}{r}.\label{bb17}
\end{equation}
At the PS radius, this becomes
\begin{equation}
    \Omega^{\rm null}_{\phi}(r_{\rm ph})=\frac{\sqrt{f(r_{\rm ph})}}{r_{\rm ph}}=\frac{(1-\alpha)^{1/2}}{R_{\rm sh}}.\label{bb18}
\end{equation}
The geodesic angular velocity depends on all spacetime parameters through $r_{\rm ph}$ and $f(r_{\rm ph})$. The inverse relationship with $R_{\rm sh}$ in Eq.~\eqref{bb18} implies that larger shadows correspond to slower photon orbital velocities, providing an additional observable signature of the CS and PFDM effects.

\section{Test Particles Dynamics around BH} \label{isec4}

In this section, we investigate the dynamics of neutral test particles around the charged Bardeen BH coupled to a CS and surrounded by PFDM. We derive the effective potential and analyze how the spacetime parameters influence the specific energy, specific angular momentum, and ISCO location for particles in circular orbits.

The dynamics of a neutral particle with mass  $m$ are described by the Hamiltonian:
\begin{equation}
    H = \frac{1}{2}g^{\mu\nu}p_{\mu}p_{\nu} + \frac{1}{2}m^2,\label{cc1}
\end{equation}
where $p^{\mu} = mu^{\mu}$ is the 4-momentum, $u^{\mu} = dx^{\mu}/d\tau$ is the 4-velocity, and $\tau$ is the proper time. The Hamiltonian equations of motion are
\begin{equation}
\frac{dx^\mu}{d\zeta} \equiv mu^\mu = \frac{\partial H}{\partial p_\mu}, \qquad 
\frac{dp_\mu}{d\zeta} = -\frac{\partial H}{\partial x^\mu},\label{cc2}
\end{equation}
where $\zeta = \tau/m$ is the affine parameter.

The static, spherically symmetric spacetime admits two Killing vectors associated with time translation and axial rotation, yielding two conserved quantities:
\begin{equation}
\frac{p_t}{m} = -f(r)\frac{dt}{d\tau} = -\mathcal{E},\label{cc3}
\end{equation}
\begin{equation}
\frac{p_\phi}{m} = r^2\sin^2\theta\frac{d\phi}{d\tau} = \mathcal{L},\label{cc4}
\end{equation}
where $\mathcal{E}$ and $\mathcal{L}$ are the specific energy and specific angular momentum (per unit mass), respectively.

The 4-velocity components are:
\begin{align}
    &\frac{dt}{d\tau} = \frac{\mathcal{E}}{f(r)},\label{cc5}\\
    &\frac{d\phi}{d\tau} = \frac{\mathcal{L}}{r^2\sin^2\theta},\label{cc6}\\
    &\frac{d\theta}{d\tau} = \frac{p_{\theta}}{mr^2},\label{cc7}\\
    &\frac{dr}{d\tau} = \sqrt{\mathcal{E}^2 - \left(\epsilon + \frac{\mathcal{L}^2}{r^2\sin^2\theta} + \frac{p^2_{\theta}}{mr^2}\right)f(r)},\label{cc8}
\end{align}
where $\epsilon = 1$ for time-like particles and $\epsilon = 0$ for light-like particles.

In our work, we focus only on time-like
particles. Therefore, for time-like particles, the Hamiltonian in Eq.~\eqref{cc1} can be re-written as:
\begin{equation}
    H = \frac{f(r)}{2}p^2_r + \frac{p^2_{\theta}}{r^2} + \frac{1}{2}\frac{m^2}{f(r)}\left[U_{\rm eff}(r,\theta) - \mathcal{E}^2\right],\label{cc9}
\end{equation}
where the effective potential $U_{\rm eff}(r,\theta)$ takes the form:
\begin{equation}
    U_{\rm eff}(r,\theta) = \left(1 + \frac{\mathcal{L}^2}{r^2\sin^2\theta}\right)f(r)=\left(1 + \frac{\mathcal{L}^2}{r^2\sin^2\theta}\right)\,\left(1 - \alpha - \frac{2Mr^2}{\left(q^2 + r^2\right)^{3/2}} + \frac{Q^2}{r^2} + \frac{\beta}{r}\ln\!\frac{r}{|\beta|}\right).\label{cc10}
\end{equation}

In the equatorial plane ($\theta = \pi/2$), the effective potential becomes:
\begin{equation}
    U_{\rm eff}(r) = \left(1 + \frac{\mathcal{L}^2}{r^2}\right)\left[1 - \alpha - \frac{2Mr^2}{(q^2+r^2)^{3/2}} + \frac{Q^2}{r^2} + \frac{\beta}{r}\ln\frac{r}{|\beta|}\right].\label{cc10b}
\end{equation}

The effective potential depends on all spacetime parameters: the CS parameter $\alpha$, the PFDM parameter $\beta$, the magnetic monopole charge $q$, the electric charge $Q$, and the BH mass $M$. Additionally, the specific angular momentum $\mathcal{L}$ modifies the centrifugal barrier contribution. The effective potential $U_{\rm eff}(r, \theta)$ is crucial for understanding the motion of the test particles, as it allows one to describe the trajectories of the particles without directly solving the equations of motion. 

\begin{figure}[ht!]
    \centering
    \includegraphics[width=0.45\linewidth]{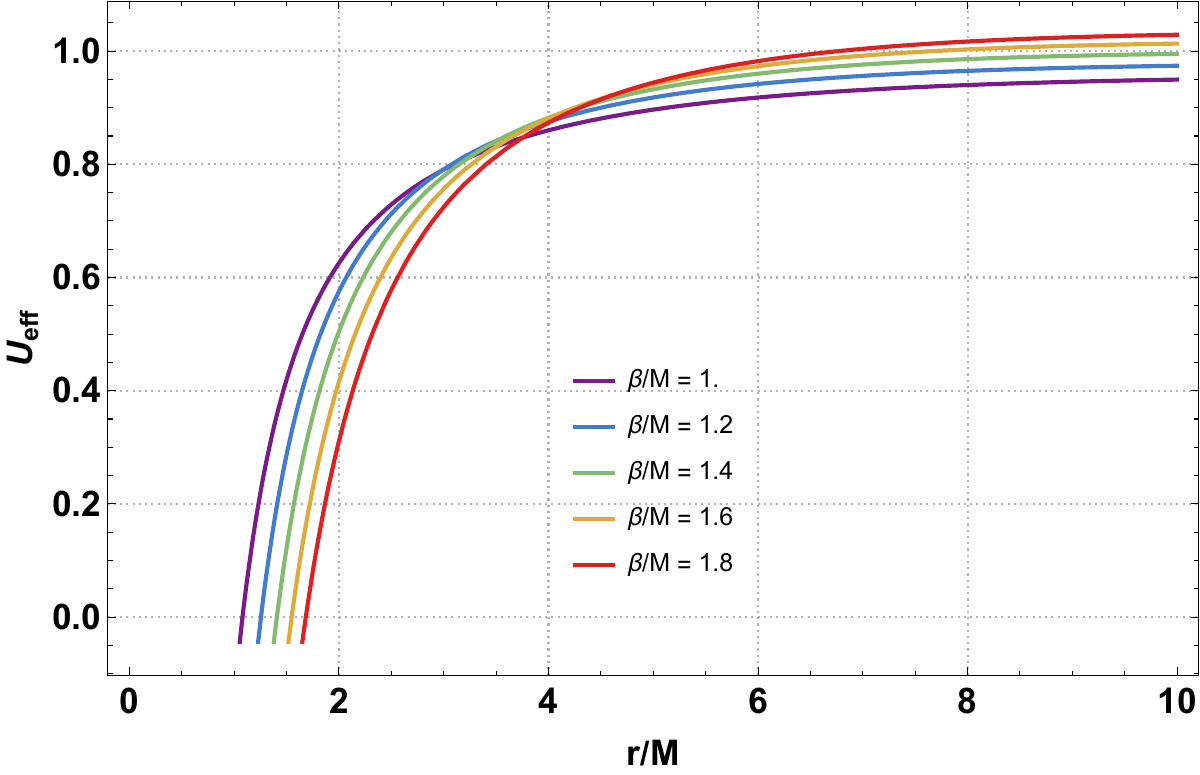}\quad
    \includegraphics[width=0.45\linewidth]{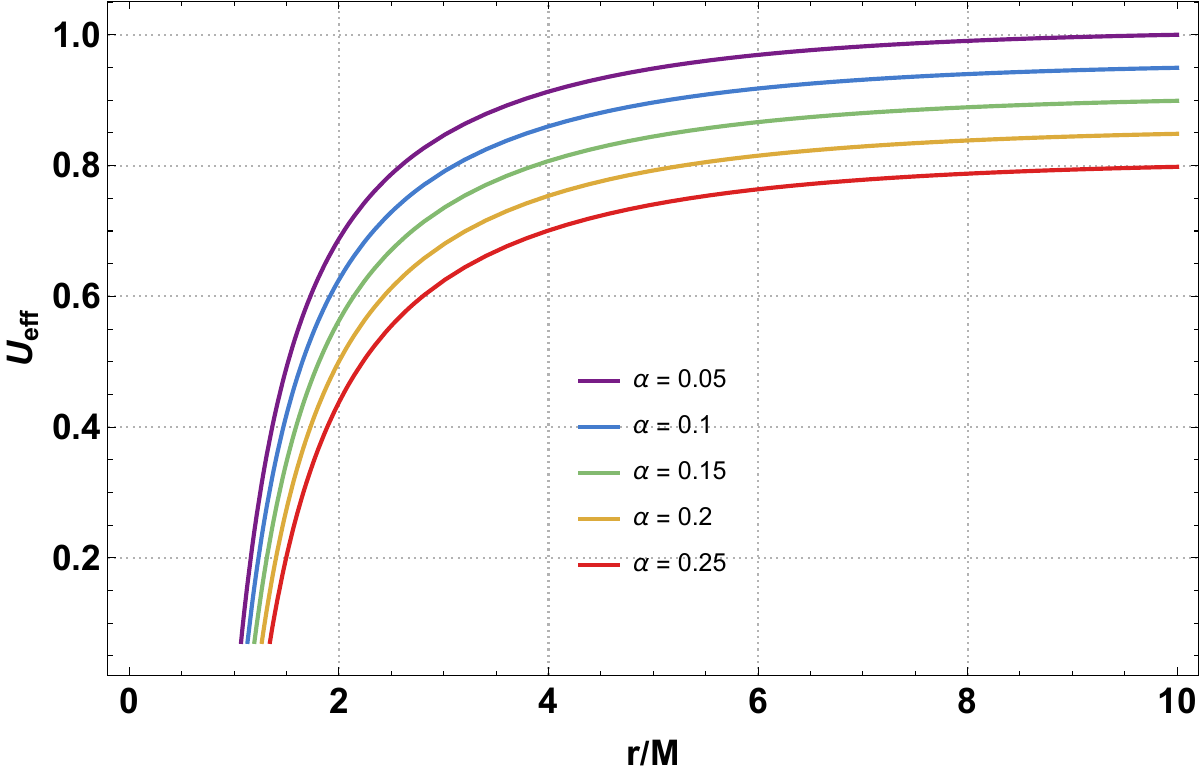}\\
    (a) $\alpha=0.1$ \hspace{6cm} (b) $\beta/M=1.0$
    \caption{Effective potential $U_{\rm eff}(r,\pi/2)$ for timelike geodesics as a function of $r/M$. Panel (a): varying $\beta/M = 1.0, 1.2, 1.4, 1.6, 1.8$ with fixed $\alpha = 0.1$. Panel (b): varying $\alpha = 0.05, 0.1, 0.15, 0.2, 0.25$ with fixed $\beta/M = 1.0$. Other parameters: $Q/M=1$, $q/M=0.1$, $\mathcal{L}/M=1$. Increasing $\beta/M$ raises the potential, while increasing $\alpha$ lowers it.}
    \label{fig:timelike}
\end{figure}

Figure~\ref{fig:timelike} displays the effective potential for various parameter combinations. In panel (a), increasing $\beta/M$ raises the potential, indicating that PFDM enhances the effective gravitational barrier for test particles. In contrast, panel (b) shows that increasing $\alpha$ lowers the potential, reducing the barrier. This opposite behavior of $\alpha$ and $\beta$ on the timelike effective potential differs from their similar effects on the null geodesic potential (cf. Fig.~\ref{fig:null}).

Next, we determine the effective force experienced by the test particles. The effective force acting on  test particles is defined as the negative gradient of the effective potential $U_{\rm eff} (r, \theta)$ and is given by 
\begin{equation}
    \mathcal{F} = -\frac{1}{2}\frac{\partial U_{\rm eff}}{\partial r}.\label{force1}
\end{equation}
Substituting the effective potential given in Eq.~\eqref{cc10} and simplifying yields:
\begin{equation}
\mathcal{F}(r,\theta) = -\frac{f'(r)}{2} + \frac{\mathcal{L}^2}{r^3\sin^2\theta}\left[2f(r) - rf'(r)\right].\label{force2}
\end{equation}

Explicitly, in the equatorial plane:
\begin{equation}
\mathcal{F}(r) = \frac{Mr(r^2-2q^2)}{(r^2+q^2)^2} + \frac{Q^2}{r^3} + \frac{\beta}{2r^2}\left(1-\ln\frac{r}{|\beta|}\right) + \frac{\mathcal{L}^2}{r^3}\left[1-\alpha-\frac{3Mr^4}{(r^2+q^2)^{5/2}}+\frac{2Q^2}{r^2}+\frac{\beta}{2r}\left(3\ln\frac{r}{|\beta|}-1\right)\right].\label{force3}
\end{equation}

\begin{figure}[ht!]
    \centering
    \includegraphics[width=0.45\linewidth]{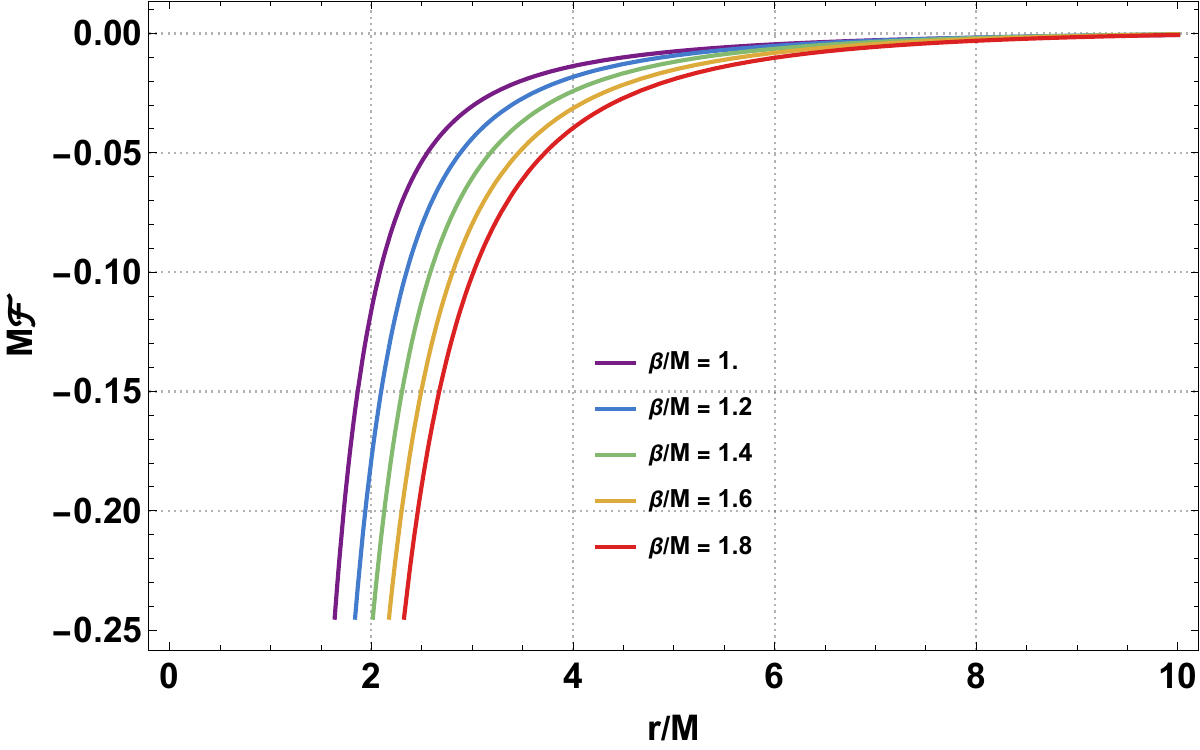}\quad
    \includegraphics[width=0.45\linewidth]{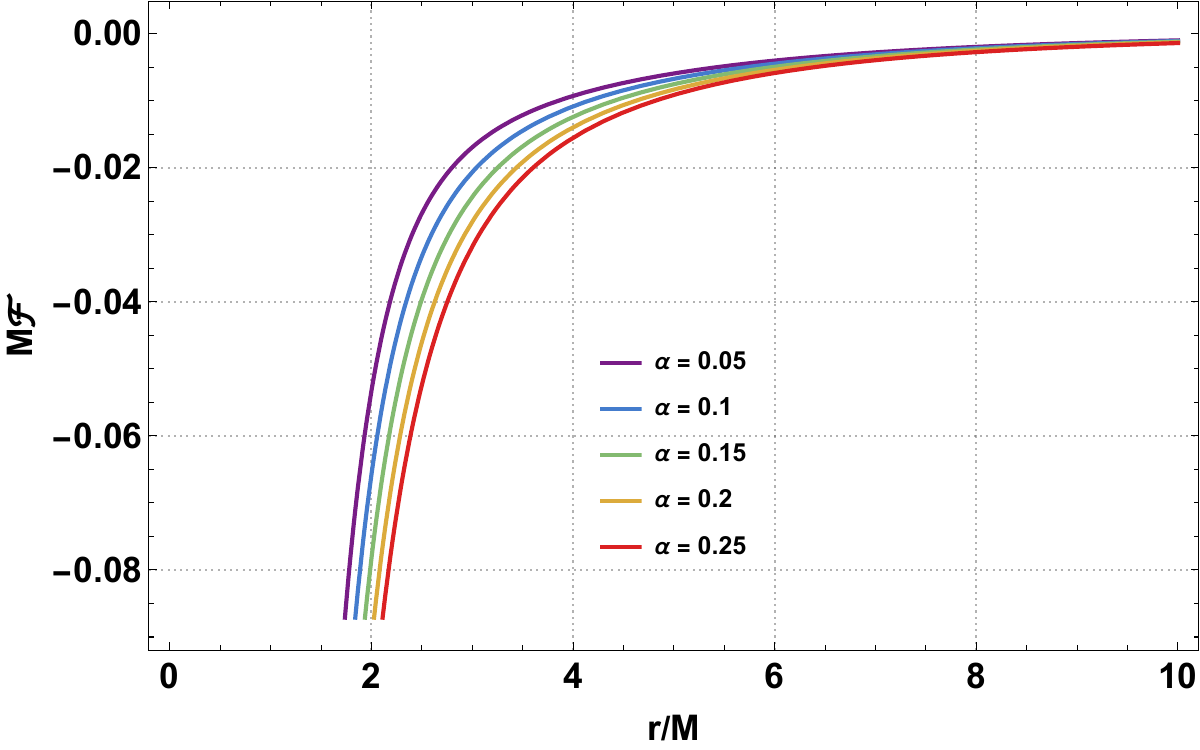}\\
    (a) $\alpha=0.1$ \hspace{6cm} (b) $\beta/M=0.8$
    \caption{Effective radial force $M\mathcal{F}(r,\pi/2)$ on test particles as a function of $r/M$. Panel (a): varying $\beta/M = 1.0, 1.2, 1.4, 1.6, 1.8$ with fixed $\alpha = 0.1$. Panel (b): varying $\alpha = 0.05, 0.1, 0.15, 0.2, 0.25$ with fixed $\beta/M = 0.8$. Other parameters: $Q/M=1$, $q/M=0.1$, $\mathcal{L}/M=1$. Both parameters reduce the force magnitude, indicating weaker gravitational binding.}
    \label{fig:force}
\end{figure}

Figure~\ref{fig:force} shows that increasing both $\beta/M$ and $\alpha$ reduces the magnitude of the effective force. This indicates that test particles experience weaker gravitational binding and are less tightly bound to the central BH for higher values of these parameters.

Circular orbits in the equatorial plane satisfy the conditions:
\begin{equation}
    U_{\rm eff}(r) = \mathcal{E}^2,\label{cc11}
\end{equation}
\begin{equation}
    \frac{\partial U_{\rm eff}}{\partial r} = 0.\label{cc12}
\end{equation}

Solving these equations yields the specific angular momentum and specific energy for circular orbits:
\begin{equation}
    \mathcal{L}^2_{\rm sp} = \frac{r^3f'(r)}{2f(r)-rf'(r)} = r^2\frac{\frac{Mr^2(r^2-2q^2)}{(q^2+r^2)^{5/2}} - \frac{Q^2}{r^2} + \frac{\beta}{2r}\left(1-\ln\frac{r}{|\beta|}\right)}{1-\alpha-\frac{3Mr^4}{(q^2+r^2)^{5/2}}+\frac{2Q^2}{r^2}+\frac{\beta}{2r}\left(3\ln\frac{r}{|\beta|}-1\right)},\label{cc13}
\end{equation}
\begin{equation}
    \mathcal{E}^2_{\rm sp} = \frac{2f^2(r)}{2f(r)-rf'(r)} = \frac{\left(1-\alpha-\frac{2Mr^2}{(q^2+r^2)^{3/2}}+\frac{Q^2}{r^2}+\frac{\beta}{r}\ln\frac{r}{|\beta|}\right)^2}{1-\alpha-\frac{3Mr^4}{(q^2+r^2)^{5/2}}+\frac{2Q^2}{r^2}+\frac{\beta}{2r}\left(3\ln\frac{r}{|\beta|}-1\right)}.\label{cc14}
\end{equation}

\begin{figure}[ht!]
    \centering
    \includegraphics[width=0.45\linewidth]{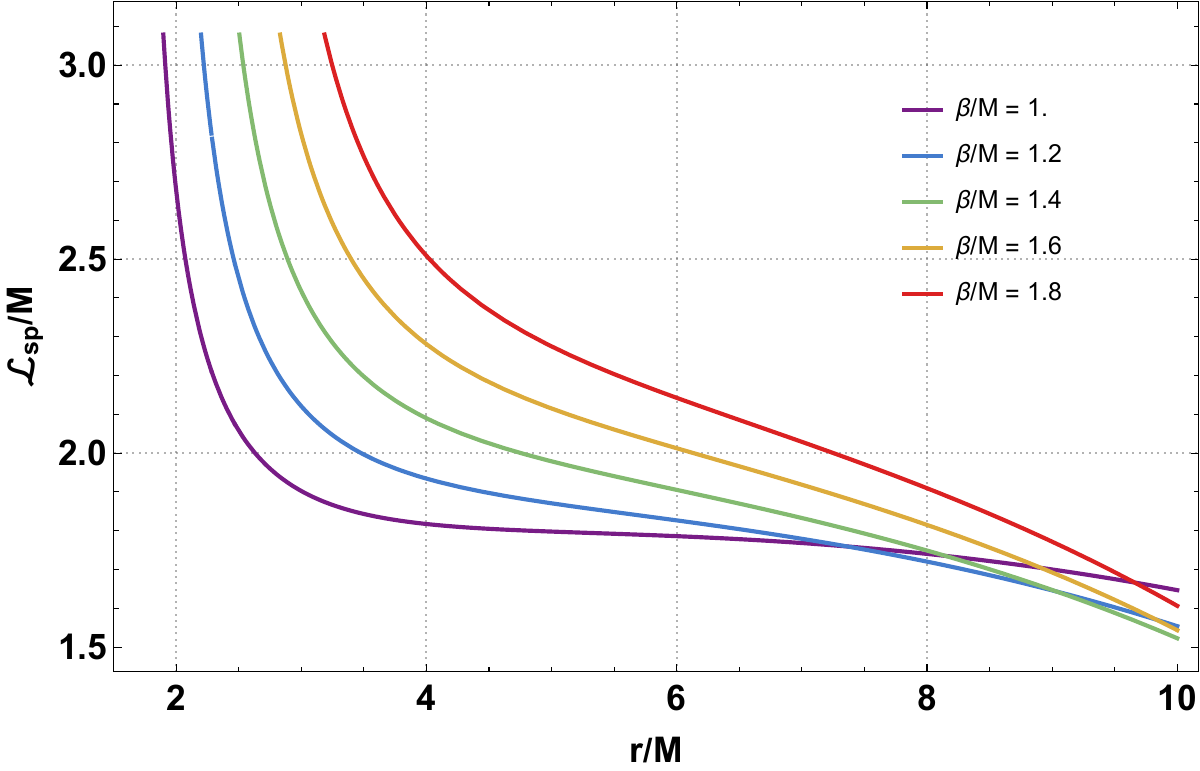}\quad
    \includegraphics[width=0.45\linewidth]{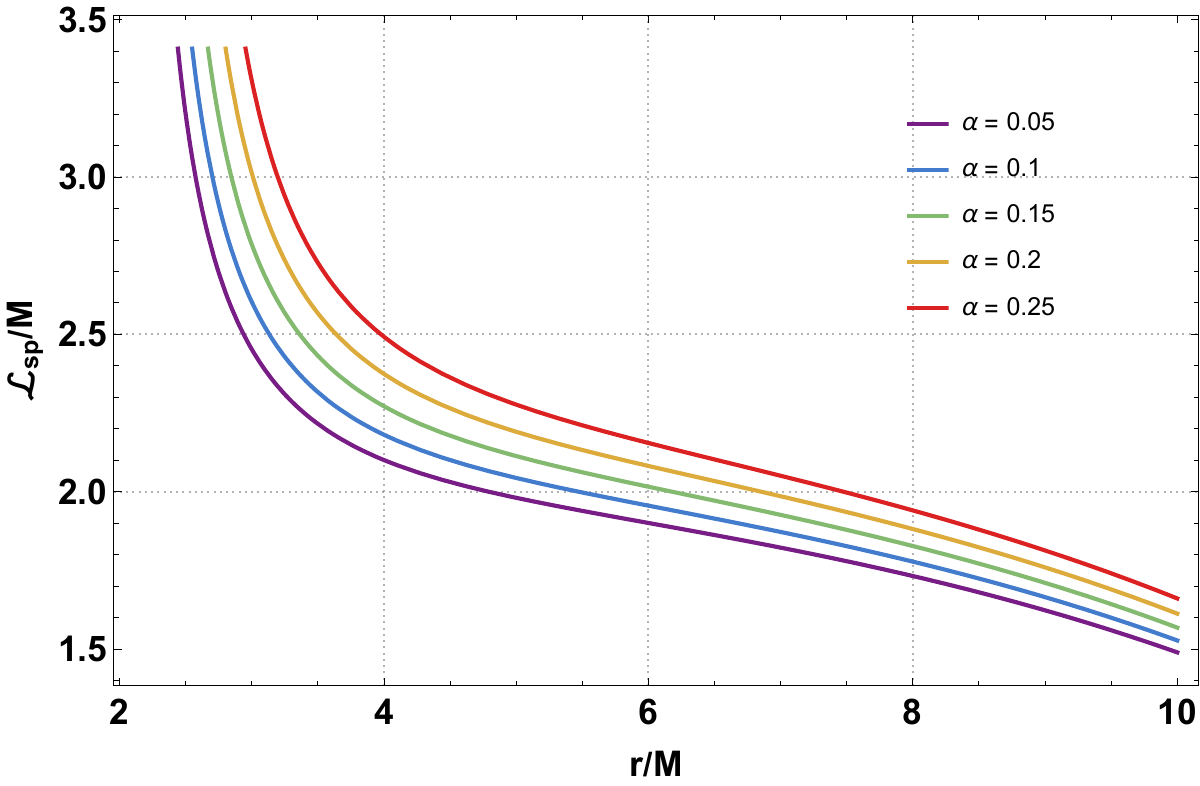}\\
    (a) $\alpha=0.1$ \hspace{6cm} (b) $\beta/M=1.5$
    \caption{Specific angular momentum $\mathcal{L}_{\rm sp}/M$ for circular orbits as a function of $r/M$. Panel (a): varying $\beta/M = 1.0, 1.2, 1.4, 1.6, 1.8$ with fixed $\alpha = 0.1$. Panel (b): varying $\alpha = 0.05, 0.1, 0.15, 0.2, 0.25$ with fixed $\beta/M = 1.5$. Other parameters: $Q/M=1$, $q/M=0.1$. Both parameters increase the required angular momentum for circular orbits.}
    \label{fig:specific-momentum}
\end{figure}

\begin{figure}[ht!]
    \centering
    \includegraphics[width=0.45\linewidth]{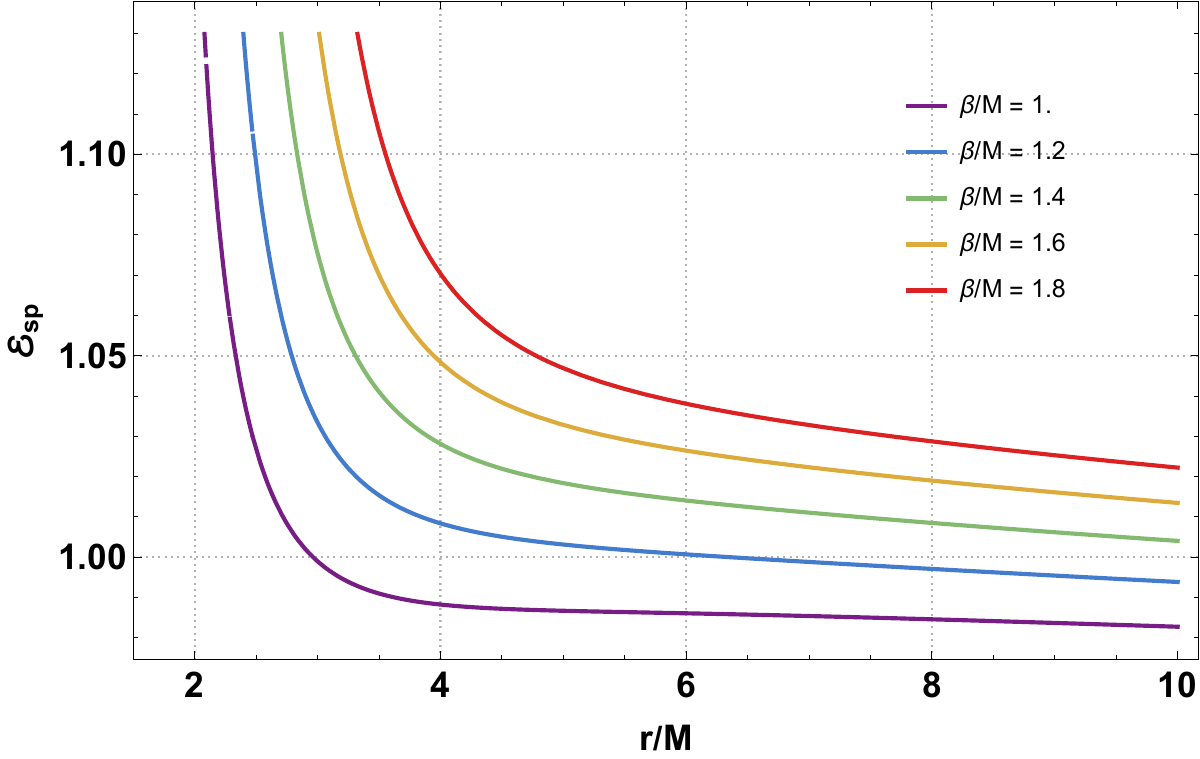}\quad
    \includegraphics[width=0.45\linewidth]{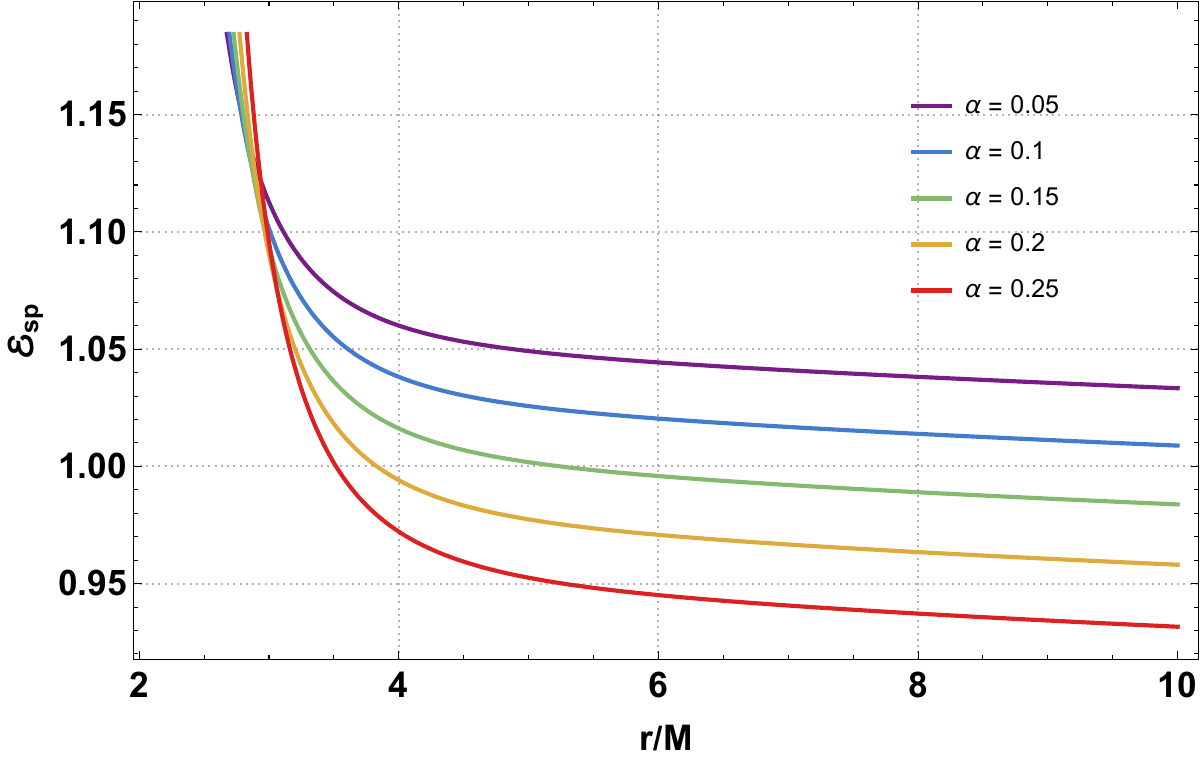}\\
    (a) $\alpha=0.1$ \hspace{6cm} (b) $\beta/M=1.5$
    \caption{Specific energy $\mathcal{E}_{\rm sp}$ for circular orbits as a function of $r/M$. Panel (a): varying $\beta/M = 1.0, 1.2, 1.4, 1.6, 1.8$ with fixed $\alpha = 0.1$. Panel (b): varying $\alpha = 0.05, 0.1, 0.15, 0.2, 0.25$ with fixed $\beta/M = 1.5$. Other parameters: $Q/M=1$, $q/M=0.1$. Increasing $\beta/M$ raises the energy, while increasing $\alpha$ lowers it at fixed $\beta/M$.}
    \label{fig:specific-energy}
\end{figure}

Figures~\ref{fig:specific-momentum} and \ref{fig:specific-energy} display the specific angular momentum and energy for circular orbits. In Fig.~\ref{fig:specific-momentum}, both $\beta/M$ and $\alpha$ increase $\mathcal{L}_{\rm sp}$, indicating that particles require higher angular momentum to maintain circular orbits in spacetimes with stronger PFDM or CS contributions.

In Fig.~\ref{fig:specific-energy}, panel (a) shows that increasing $\beta/M$ raises $\mathcal{E}_{\rm sp}$, implying that more energy is required for circular orbits in PFDM-rich environments. Panel (b) demonstrates the opposite trend: increasing $\alpha$ at fixed $\beta/M$ reduces $\mathcal{E}_{\rm sp}$, indicating that the CS effect lowers the energy requirement for bound orbits.

The stability of circular orbits is determined by the second derivative of the effective potential. Stable orbits correspond to local minima ($\partial^2 U_{\rm eff}/\partial r^2 > 0$), while unstable orbits correspond to local maxima.  In Newtonian theory, the effective potential always has a minimum for any given angular momentum, and there are no ISCOs with a minimum
radius. However, when the form of the effective potential depends on the angular momentum of the particle and other parameters, the situation changes. The ISCO marks the boundary between stable and unstable circular orbits, defined by the conditions:
\begin{equation}
U_{\rm eff}(r) = \mathcal{E}^2, \qquad \frac{\partial U_{\rm eff}}{\partial r} = 0, \qquad \frac{\partial^2 U_{\rm eff}}{\partial r^2} = 0.\label{cc25}
\end{equation}

Using Eq.~\eqref{cc10}, the ISCO condition reduces to:
\begin{equation}
    \frac{3}{r}f(r)f'(r) - 2[f'(r)]^2 + f(r)f''(r) = 0.\label{cc26}
\end{equation}

For a Schwarzschild BH, the ISCO is located at $r_{\rm ISCO} = 6M = 3r_g$, where $r_g = 2M$ is the Schwarzschild radius. The presence of the CS and PFDM parameters, along with the magnetic and electric charges, modifies the ISCO location. Due to the complexity of Eq.~\eqref{cc26} with all parameters included, the ISCO radius must be determined numerically.

Table~\ref{tab:particle_summary} summarizes how the CS and PFDM parameters affect the dynamics of neutral test particles. The results reveal that these two parameters do not act uniformly across all dynamical quantities, producing a characteristic pattern that distinguishes their individual contributions to the spacetime geometry.

\begin{table}[ht!]
\centering
\setlength{\tabcolsep}{10pt}
\renewcommand{\arraystretch}{1.5}
\begin{tabular}{|l|c|c|p{5.0cm}|}
\hline
\rowcolor{orange!50}
\textbf{Quantity} & \textbf{Effect of $\alpha \uparrow$} & \textbf{Effect of $\beta \uparrow$} & \textbf{Physical Implication} \\
\hline
Effective potential $U_{\rm eff}$ & Decreases & Increases & Opposite effects on gravitational barrier height \\
\hline
Radial force $|\mathcal{F}|$ & Decreases & Decreases & Weaker gravitational binding to the central BH \\
\hline
Specific angular momentum $\mathcal{L}_{\rm sp}$ & Increases & Increases & Higher rotation required for circular orbits \\
\hline
Specific energy $\mathcal{E}_{\rm sp}$ & Decreases & Increases & Opposite effects on orbital binding energy \\
\hline
ISCO radius $r_{\rm ISCO}$ & Increases & Increases & Outward shift of the innermost stable orbit \\
\hline
\end{tabular}
\caption{Effects of CS parameter $\alpha$ and PFDM parameter $\beta$ on neutral particle dynamics in the equatorial plane. The opposite responses of $U_{\rm eff}$ and $\mathcal{E}_{\rm sp}$ to these parameters, combined with their common influence on $r_{\rm ISCO}$, create a distinctive dynamical signature.}
\label{tab:particle_summary}
\end{table}

A particularly notable feature in Table~\ref{tab:particle_summary} is the opposite behavior of the effective potential $U_{\rm eff}$ and specific energy $\mathcal{E}_{\rm sp}$ under variations of $\alpha$ and $\beta$: the CS parameter lowers both quantities, while PFDM raises both. This dichotomy originates from the different functional forms through which these parameters enter the metric—$\alpha$ as a constant deficit $(1-\alpha)$ and $\beta$ through the logarithmic term $(\beta/r)\ln(r/\beta)$. Despite these opposite effects on $U_{\rm eff}$ and $\mathcal{E}_{\rm sp}$, both parameters shift the ISCO outward, indicating that stable circular orbits exist only at larger radii when CS or PFDM contributions are stronger. The outward displacement of the ISCO directly affects the inner edge of accretion disks around BHs, which in turn modifies the thermal continuum spectrum and characteristic frequencies observed in X-ray binary systems \cite{sec4is01,sec4is02}.

\section{Quasi-periodic oscillations (QPO)\lowercase{s}} \label{isec5}

QPOs are nearly periodic variations observed in the X-ray flux from accreting compact objects such as BHs and neutron stars. Unlike strictly periodic signals, QPOs manifest as sharp peaks with finite widths in the power spectrum, indicating slight variations or modulations in frequency over time \cite{sec5is01}. QPOs originate from several physical mechanisms:
\begin{itemize}
    \item Oscillations of matter in the inner regions of the accretion disk surrounding the compact object.
    \item Orbital motion, disk instabilities, or resonances occurring in the strong gravitational field near the compact object.
\end{itemize}

Two main classes of QPOs are distinguished:
\begin{itemize}
    \item \textbf{LFQPOs:} Typically observed at frequencies of a few Hertz, these are commonly associated with disk instabilities or Lense-Thirring precession phenomena.
    \item \textbf{HFQPOs:} Observed at frequencies of hundreds of Hertz, especially in stellar-mass BHs, and often linked to relativistic orbital frequencies near the ISCO.
\end{itemize}

QPOs provide a valuable probe of strong-field gravity near BHs and neutron stars. Furthermore, they exhibit potential connections to QNMs, offering observational windows into the fundamental properties of compact objects and their surrounding spacetime \cite{isz35,isz37}. To study the oscillatory motion of neutral test particles, we perturb the equations of motion around stable circular orbits. A test particle slightly displaced from its equilibrium position in a stable circular orbit experiences epicyclic motion with linear harmonic oscillations.

The Keplerian frequency in physical units is given by \cite{Stella1,Stella2,Stuchlik}
\begin{equation}
   \nu_{K,r,\theta} = \frac{1}{2\pi}\frac{c^3}{GM}\,\Omega_{r,\theta,\phi}\quad [\mathrm{Hz}].\label{pp1}
\end{equation}
The azimuthal frequency $\Omega_{\phi}$, representing the frequency of Keplerian orbits in the $\phi$-direction, is
\begin{equation}
   \Omega_{\phi}=\frac{d\phi}{dt}=\frac{\dot{\phi}}{\dot{t}}=\frac{f'(r)}{2r}=\frac{1}{r^2}\left[-Mr^2\frac{2q^2-r^2}{(q^2+r^2)^{5/2}}-\frac{Q^2}{r^2}+\frac{\beta}{2r}\left(1-\ln\frac{r}{|\beta|}\right)\right].\label{pp2}
\end{equation}

From Eq.~\eqref{pp2}, the azimuthal frequency depends explicitly on the PFDM parameter $\beta$, the magnetic monopole charge $q$, the electric charge $Q$, and the BH mass $M$. Notably, the CS parameter $\alpha$ does not appear in $\Omega_\phi$, indicating that the string cloud has no direct effect on the Keplerian orbital frequency.

\begin{figure}[ht!]
    \centering
    \includegraphics[width=0.55\linewidth]{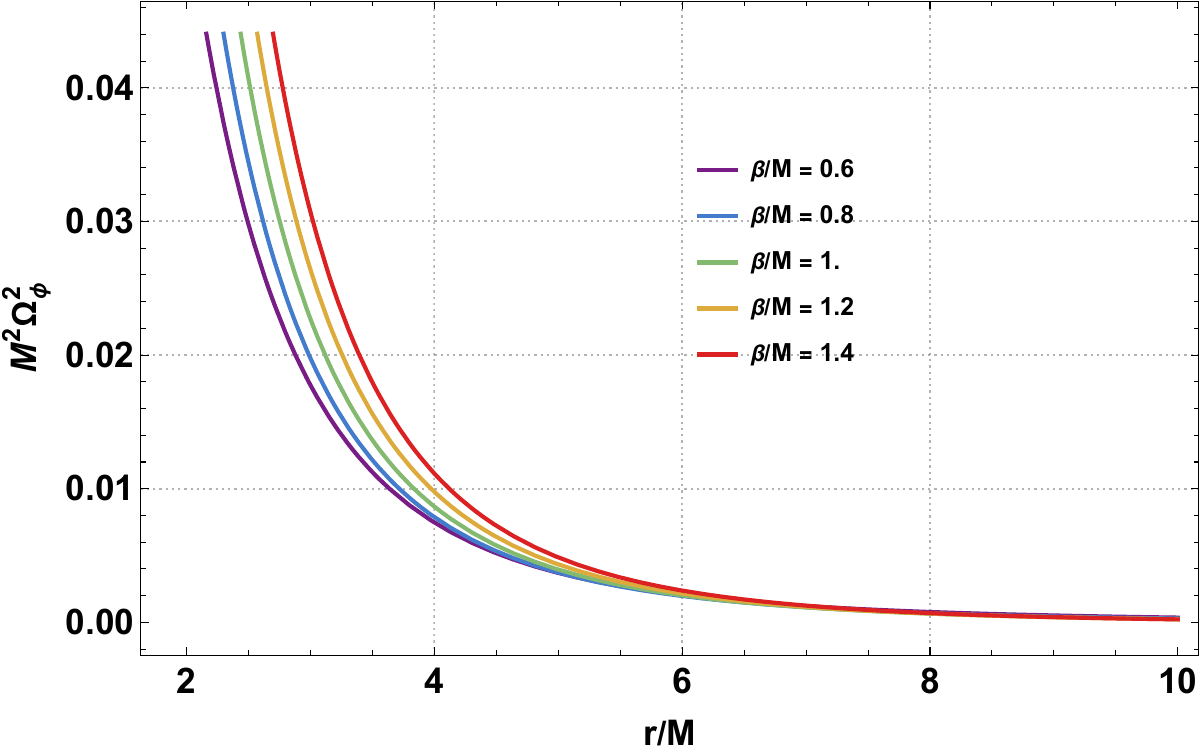}
    \caption{Azimuthal frequency $M^2\Omega_{\phi}^2$ as a function of $r/M$ for various $\beta/M = 0.6, 0.8, 1.0, 1.2, 1.4$ with fixed $Q/M=1$, $q/M=0.1$. The frequency increases with $\beta/M$, demonstrating the significant influence of PFDM on Keplerian orbital dynamics.}
    \label{fig:azimuthal-frequency}
\end{figure}

Figure~\ref{fig:azimuthal-frequency} displays $\Omega_\phi$ as a function of $r/M$ for various values of $\beta/M$. Increasing $\beta/M$ leads to higher azimuthal frequencies, indicating that PFDM enhances the orbital angular velocity of test particles. This effect becomes more pronounced at smaller radii where the dark matter influence is stronger.

The radial ($\nu_r = \Omega_r/2\pi M$) and vertical ($\nu_\theta = \Omega_\theta/2\pi M$) epicyclic frequencies characterize small oscillations around stable circular orbits. These frequencies are obtained from the second derivatives of the effective potential $U_{\rm eff}(r,\theta)$ in Eq.~\eqref{cc10} and are critical for analyzing QPO phenomena in BHs \cite{isz37}.

Test particles in stable equatorial orbits undergo harmonic oscillations due to small displacements $\delta r$ and $\delta\theta$ from equilibrium. The oscillation equations are
\begin{equation}
\frac{d^2\delta r}{dt^2}+\Omega_r^2\,\delta r=0,\qquad \frac{d^2\delta\theta}{dt^2}+\Omega_\theta^2\,\delta\theta=0,\label{original}
\end{equation}
where the epicyclic frequencies as measured by a distant static observer are \cite{Stella1,Stella2,Stuchlik,Kolos2017}
\begin{equation}
\Omega_r^2=-\frac{1}{2g_{rr}(u^t)^2}\frac{\partial^2 U_{\rm eff}}{\partial r^2}\bigg|_{\theta=\pi/2}=-\frac{1}{2}\left[\frac{3}{r}f(r)f'(r)-2(f'(r))^2+f(r)f''(r)\right]f(r),\label{pp3}
\end{equation}
And 
\begin{equation}
\Omega_\theta^2=-\frac{1}{2g_{\theta\theta}(u^t)^2}\frac{\partial^2 U_{\rm eff}}{\partial\theta^2}\bigg|_{\theta=\pi/2}=-\frac{f(r)f'(r)}{2r}.\label{pp4}
\end{equation}

The radial profiles of the orbital and epicyclic frequencies of the neutral test particle motion in the field of charged Bardeen black holes with CS and PFDM have several properties fundamentally different in comparison with the radial profiles of the frequencies characterizing the geodesic epicyclic motion. At the ISCO radius $r = r_{\rm ISCO}$ satisfying Eq.~\eqref{cc26}, the radial epicyclic frequency vanishes: $\Omega_r(r_{\rm ISCO}) = 0$. This condition marks the boundary between stable and unstable circular orbits.

\begin{figure}[ht!]
    \centering
    \includegraphics[width=0.45\linewidth]{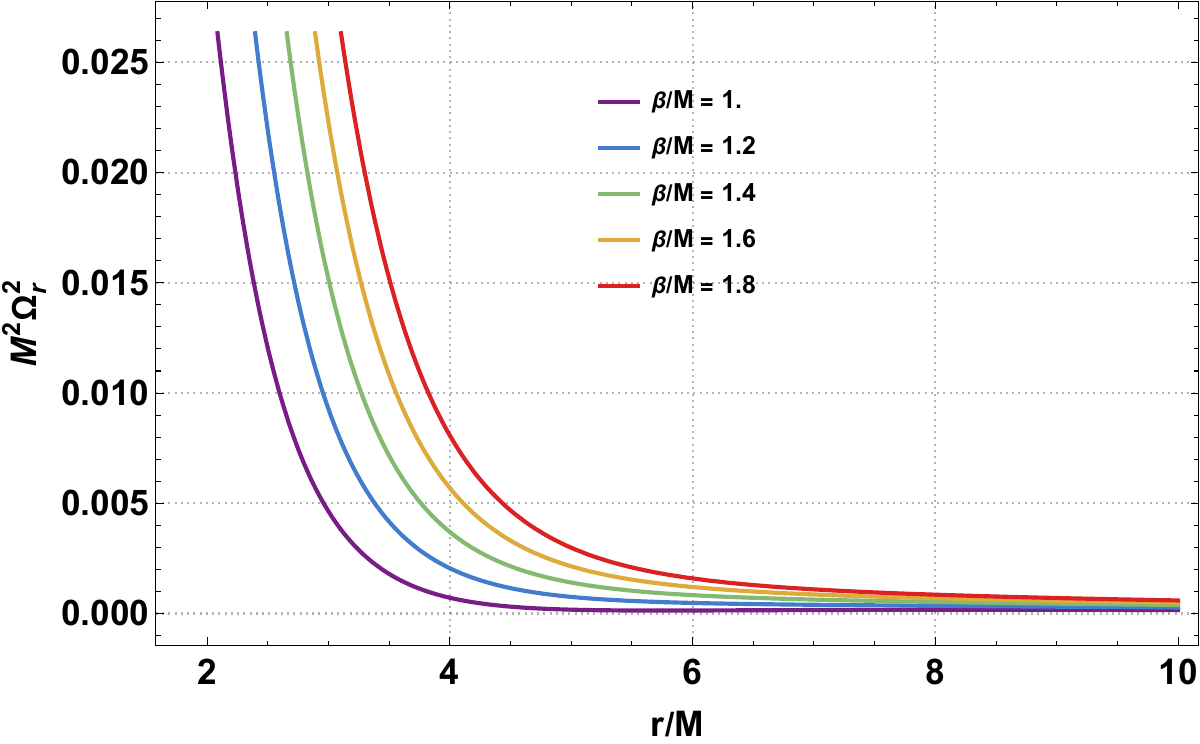}\quad
    \includegraphics[width=0.45\linewidth]{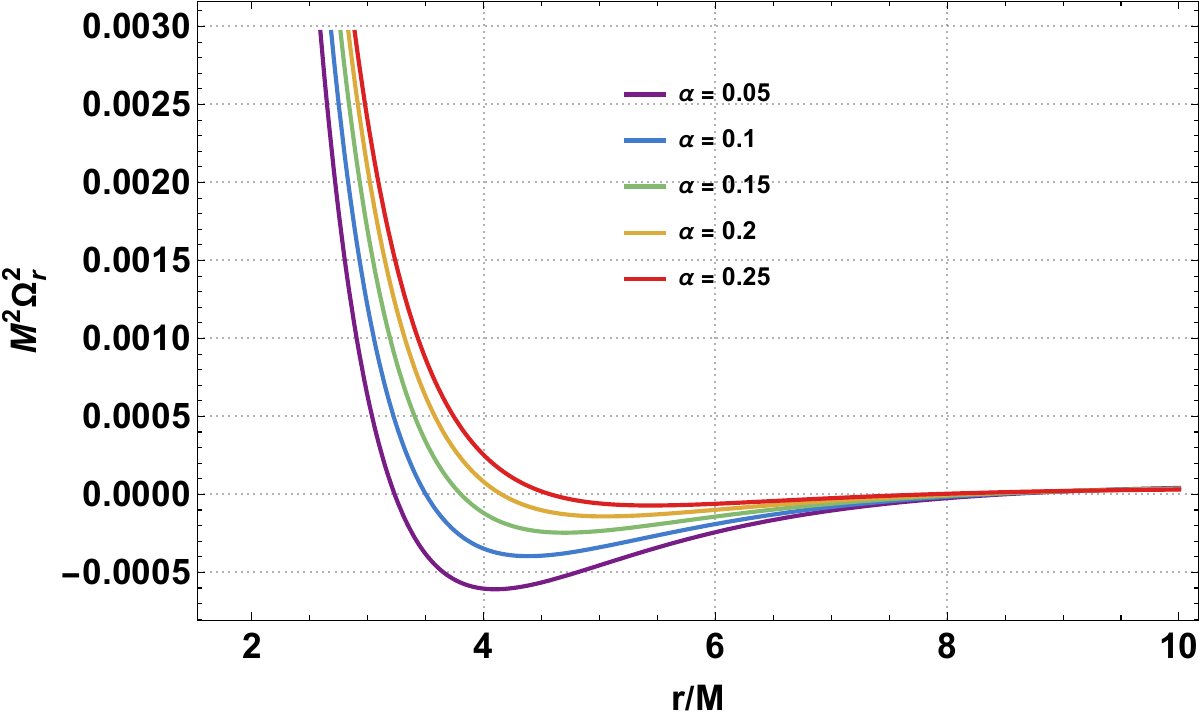}\\
    (a) $\alpha=0.1$ \hspace{6cm} (b) $\beta/M=0.8$
    \caption{Radial epicyclic frequency $M^2\Omega_r^2$ as a function of $r/M$. Panel (a): varying $\beta/M = 1.0, 1.2, 1.4, 1.6, 1.8$ with fixed $\alpha = 0.1$. Panel (b): varying $\alpha = 0.05, 0.1, 0.15, 0.2, 0.25$ with fixed $\beta/M = 0.8$. Other parameters: $Q/M=1$, $q/M=0.1$. Both $\beta/M$ and $\alpha$ increase the radial frequency, enhancing radial oscillations.}
    \label{fig:radial-frequency}
\end{figure}

\begin{figure}[ht!]
    \centering
    \includegraphics[width=0.45\linewidth]{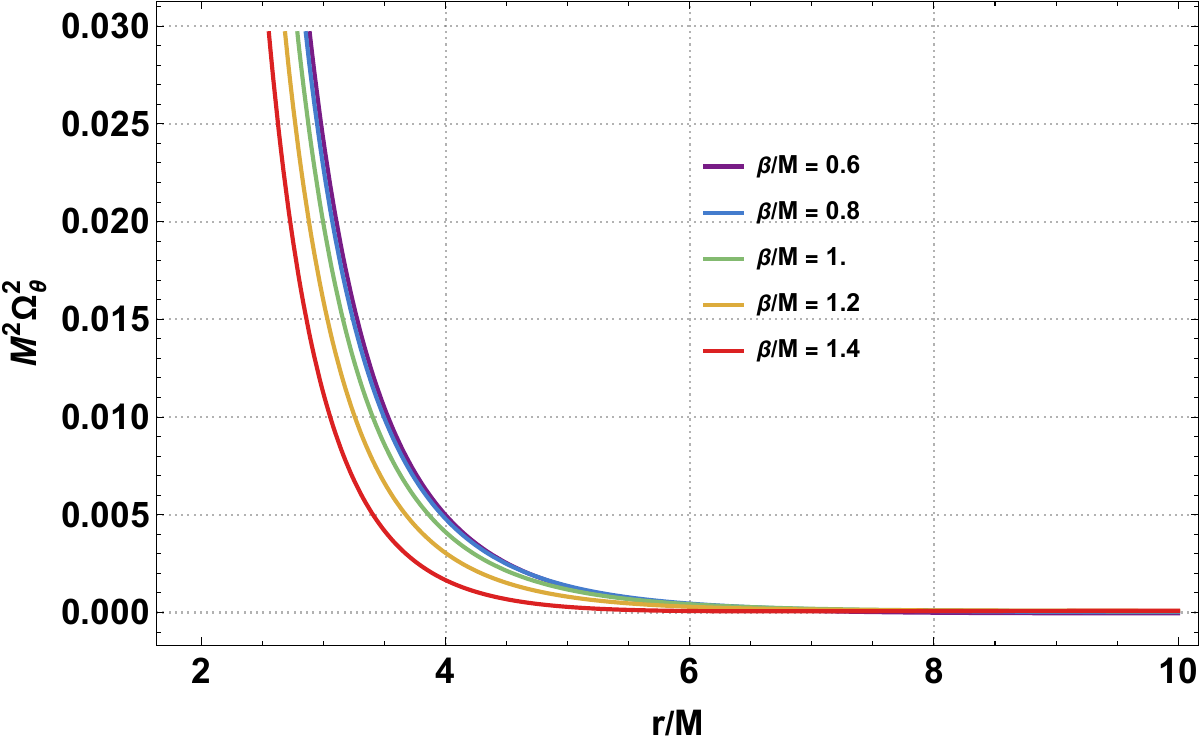}\quad
    \includegraphics[width=0.45\linewidth]{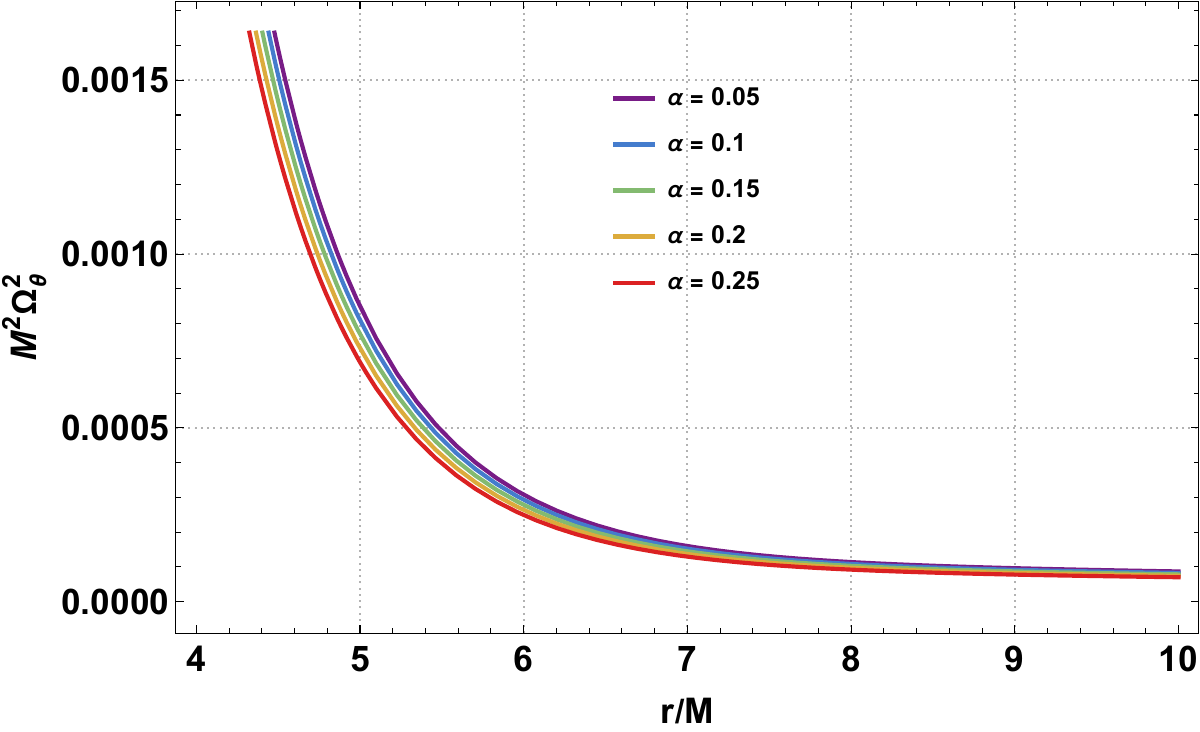}\\
    (a) $\alpha=0.1$ \hspace{6cm} (b) $\beta/M=1.2$
    \caption{Vertical epicyclic frequency $M^2\Omega_\theta^2$ as a function of $r/M$. Panel (a): varying $\beta/M = 0.6, 0.8, 1.0, 1.2, 1.4$ with fixed $\alpha = 0.1$. Panel (b): varying $\alpha = 0.05, 0.1, 0.15, 0.2, 0.25$ with fixed $\beta/M = 1.2$. Other parameters: $Q/M=2$, $q/M=0.5$. In contrast to the radial case, both parameters decrease the vertical frequency.}
    \label{fig:vertical-frequency}
\end{figure}

Figure~\ref{fig:radial-frequency} shows the radial epicyclic frequency dependence on $\beta/M$ and $\alpha$. Increasing either parameter enhances $\Omega_r^2$, indicating stronger radial oscillations. This behavior suggests that PFDM and CS destabilize circular orbits by increasing the restoring force for radial perturbations.

Figure~\ref{fig:vertical-frequency} displays the vertical epicyclic frequency. In contrast to the radial case, increasing $\beta/M$ and $\alpha$ decreases $\Omega_\theta^2$, implying weaker vertical oscillations. This asymmetry between radial and vertical responses to the spacetime parameters may produce distinctive QPO frequency ratios observable in X-ray binaries.

For a local observer comoving with the orbiting particle, the epicyclic frequencies are \cite{Stella1,Stella2,Stuchlik,Kolos2017}
\begin{equation}
   \omega_r^2=-\frac{1}{2g_{rr}}\frac{\partial^2 U_{\rm eff}}{\partial r^2},\label{pp5}
\end{equation}
\begin{equation}
   \omega_\theta^2=-\frac{1}{2g_{\theta\theta}}\frac{\partial^2 U_{\rm eff}}{\partial\theta^2}.\label{pp6}
\end{equation}
The relationship between local and distant observer frequencies is $\Omega_i^2 = \omega_i^2/(u^t)^2$, where
\begin{equation}
u^t = \dot{t} = \sqrt{\frac{2}{2f(r)-rf'(r)}}
\end{equation}
accounts for the gravitational time dilation.

For a test particle in a slightly eccentric orbit around a non-rotating BH with CS and PFDM, the periastron precession frequency $\Omega_p$ quantifies the rate at which the orbital perihelion advances. It is defined as the difference between the azimuthal and radial frequencies:
\begin{equation}
  \Omega_p = \Omega_{\phi} - \Omega_r = \frac{f'(r)}{2r}-\sqrt{-\frac{1}{2}\left[\frac{3}{r}f(r)f'(r)-2(f'(r))^2+f(r)f''(r)\right]f(r)}.\label{pp7}
\end{equation}

\begin{figure}[ht!]
    \centering
    \includegraphics[width=0.65\linewidth]{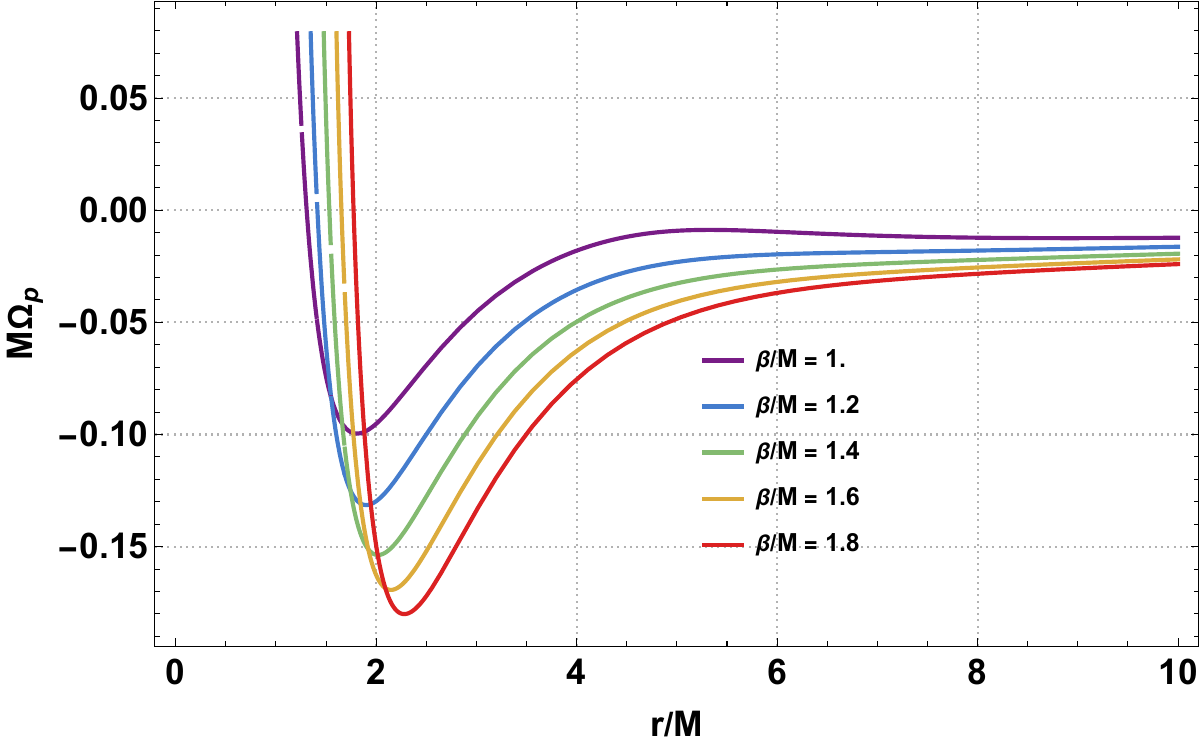}
    \caption{Periastron precession frequency $M\Omega_p$ as a function of $r/M$ for various $\beta/M = 1.0, 1.2, 1.4, 1.6, 1.8$ with fixed $Q/M=1$, $q/M=0.1$. Increasing $\beta/M$ decreases the precession frequency, indicating slower orbital precession in PFDM-rich environments.}
    \label{fig:periastron-frequency}
\end{figure}

Figure~\ref{fig:periastron-frequency} shows $\Omega_p$ as a function of $r/M$ for various PFDM parameters. Increasing $\beta/M$ decreases the periastron frequency, similar to the vertical frequency behavior. Since $\Omega_\phi$ is independent of the CS parameter $\alpha$ while $\Omega_r$ depends on it, the periastron precession provides a means to disentangle the effects of PFDM and CS through combined observations of multiple QPO frequencies.

Table~\ref{tab:qpo_summary} summarizes how the CS and PFDM parameters affect the various frequencies relevant to QPO analysis.

\begin{table}[ht!]
\centering
\setlength{\tabcolsep}{10pt}
\renewcommand{\arraystretch}{1.5}
\begin{tabular}{|l|c|c|p{5.0cm}|}
\hline
\rowcolor{orange!50}
\textbf{Frequency} & \textbf{Effect of $\alpha \uparrow$} & \textbf{Effect of $\beta \uparrow$} & \textbf{Physical Implication} \\
\hline
Azimuthal $\Omega_\phi$ & No effect & Increases & Faster Keplerian orbits; PFDM enhances orbital velocity \\
\hline
Radial $\Omega_r$ & Increases & Increases & Enhanced radial oscillations; stronger restoring force \\
\hline
Vertical $\Omega_\theta$ & Decreases & Decreases & Weaker vertical stability; reduced confinement to equatorial plane \\
\hline
Periastron $\Omega_p$ & Indirect & Decreases & Slower orbital precession; modified apsidal advance rate \\
\hline
\end{tabular}
\caption{Dependence of QPO-related frequencies on the CS parameter $\alpha$ and PFDM parameter $\beta$. The azimuthal frequency $\Omega_\phi$ is independent of $\alpha$, while radial and vertical frequencies exhibit opposite trends under parameter variations.}
\label{tab:qpo_summary}
\end{table}

Several key features emerge from Table~\ref{tab:qpo_summary}. First, the azimuthal frequency $\Omega_\phi$ depends only on $\beta$, $q$, and $Q$, but is completely independent of the CS parameter $\alpha$, as evident from Eq.~\eqref{pp2} where no $\alpha$ term appears. This independence allows, in principle, the isolation of PFDM effects from CS contributions. Second, the radial and vertical epicyclic frequencies exhibit opposite trends—both $\alpha$ and $\beta$ increase $\Omega_r$ while decreasing $\Omega_\theta$. This asymmetry modifies the frequency ratios $\Omega_r/\Omega_\phi$ and $\Omega_\theta/\Omega_\phi$ in ways distinct from standard Kerr BH predictions \cite{sec5is03,sec5is04}. Third, the periastron precession frequency $\Omega_p = \Omega_\phi - \Omega_r$ receives an indirect contribution from $\alpha$ through $\Omega_r$, and decreases with increasing $\beta$, indicating slower orbital precession in PFDM-rich environments. These distinct frequency responses to the spacetime parameters could serve as theoretical benchmarks for future precision timing studies of accreting BH systems \cite{sec5is03,sec5is04}.

\section{Scalar Perturbations and Grey-body Factors} \label{isec6}

Scalar perturbations of BHs offer a powerful probe of spacetime stability and dynamical properties under external disturbances. When a BH is surrounded by PFDM and coupled to a CS, the background geometry deviates from standard vacuum solutions, altering the behavior of scalar fields propagating in this spacetime. The scalar field evolution is governed by the Klein-Gordon equation on the curved background, which reduces to a Schrödinger-like wave equation with an effective potential depending explicitly on the metric function and hence on the PFDM and CS parameters \cite{isz38,Becar2023,QT2025}.

The effective potential determines the QNM spectrum—complex frequencies characterizing the oscillation and decay of perturbations-which encodes information about the underlying gravitational field and its coupling to exotic matter components. Studies of BHs immersed in PFDM report that the height and width of the effective potential barrier change with the dark matter density parameter, modifying both the real part (oscillation frequency) and imaginary part (damping rate) of QNMs \cite{Becar2023,QT2025}. Such effects may produce observable imprints in the ringdown phase of GWs emitted during BH mergers.

When a CS is included, its contributions similarly influence the QNM spectrum. Investigations demonstrate that the CS parameter alters the effective potential, affecting QNM frequencies and decay rates \cite{YL2023}. Importantly, the imaginary part of the quasinormal frequencies remains negative, signifying stability against linear scalar perturbations. The combined effects of PFDM and CS may lead to rich structures in the perturbative dynamics, as each component contributes differently to the effective potential profile.

The dynamics of a massless scalar field $\Phi$ in the BH background is governed by the Klein-Gordon equation:
\begin{equation}
\frac{1}{\sqrt{-g}}\partial_\mu\left(\sqrt{-g}\,g^{\mu\nu}\partial_\nu\Phi\right) = 0.\label{ss1}
\end{equation}

Using the separation ansatz $\Phi = \frac{\psi(r)}{r}Y_{\ell m}(\theta,\phi)\,e^{-i\omega t}$, where $Y_{\ell m}$ are the spherical harmonics and $\omega$ is the QNM frequency, the Klein-Gordon equation reduces to a Schrödinger-like wave equation:
\begin{equation}
\frac{d^2\psi}{dr_*^2} + \left[\omega^2 - V_{\rm s}(r)\right]\psi = 0,\label{ss2}
\end{equation}
where $dr_* = dr/f(r)$ defines the tortoise coordinate satisfying $r_* \to -\infty$ at the horizon and $r_* \to +\infty$ at spatial infinity. The effective potential for scalar perturbations is
\begin{equation}
V_{\rm s}(r) = \frac{f(r)}{r^2}\left[\ell(\ell+1)+rf'(r)\right],\label{ss3}
\end{equation}
where $\ell = 0, 1, 2, \ldots$ is the multipole number.

Substituting the metric function from Eq.~\eqref{function} yields the explicit form:
\begin{equation}
V_{\rm s}(r) = \frac{1 - \alpha - \frac{2Mr^2}{(r^2+q^2)^{3/2}} + \frac{Q^2}{r^2} + \frac{\beta}{r}\ln\frac{r}{\beta}}{r^2}\left[\ell(\ell+1) + \frac{2Mr^2(r^2-2q^2)}{(r^2+q^2)^{5/2}} - \frac{2Q^2}{r^2} + \frac{\beta}{r}\left(1-\ln\frac{r}{\beta}\right)\right].\label{ss4}
\end{equation}

The potential in Eq.~\eqref{ss4} depends on all five spacetime parameters: the CS parameter $\alpha$, the PFDM parameter $\beta$, the magnetic monopole charge $q$, the electric charge $Q$, and the BH mass $M$. Each parameter influences the spacetime curvature and modifies the effective barrier experienced by scalar waves, affecting the height, shape, and location of the potential peak.

\begin{figure}[ht!]
    \centering
    \includegraphics[width=0.45\linewidth]{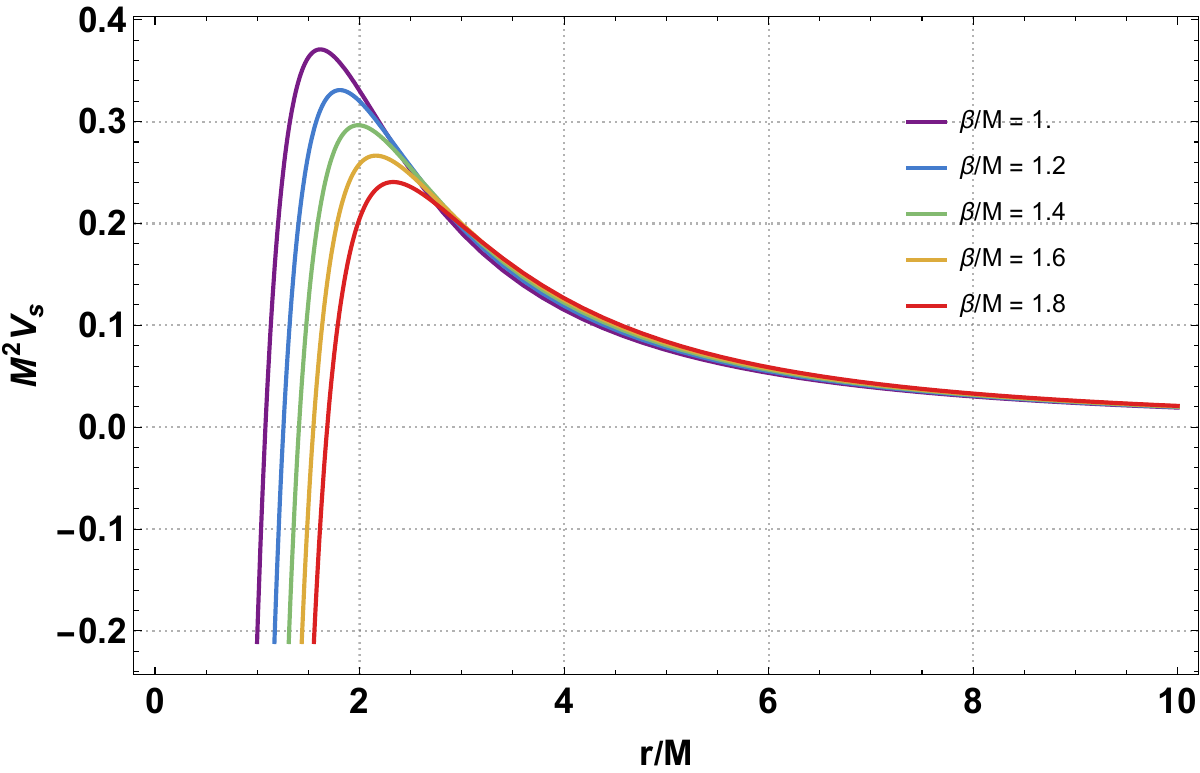}\quad
    \includegraphics[width=0.45\linewidth]{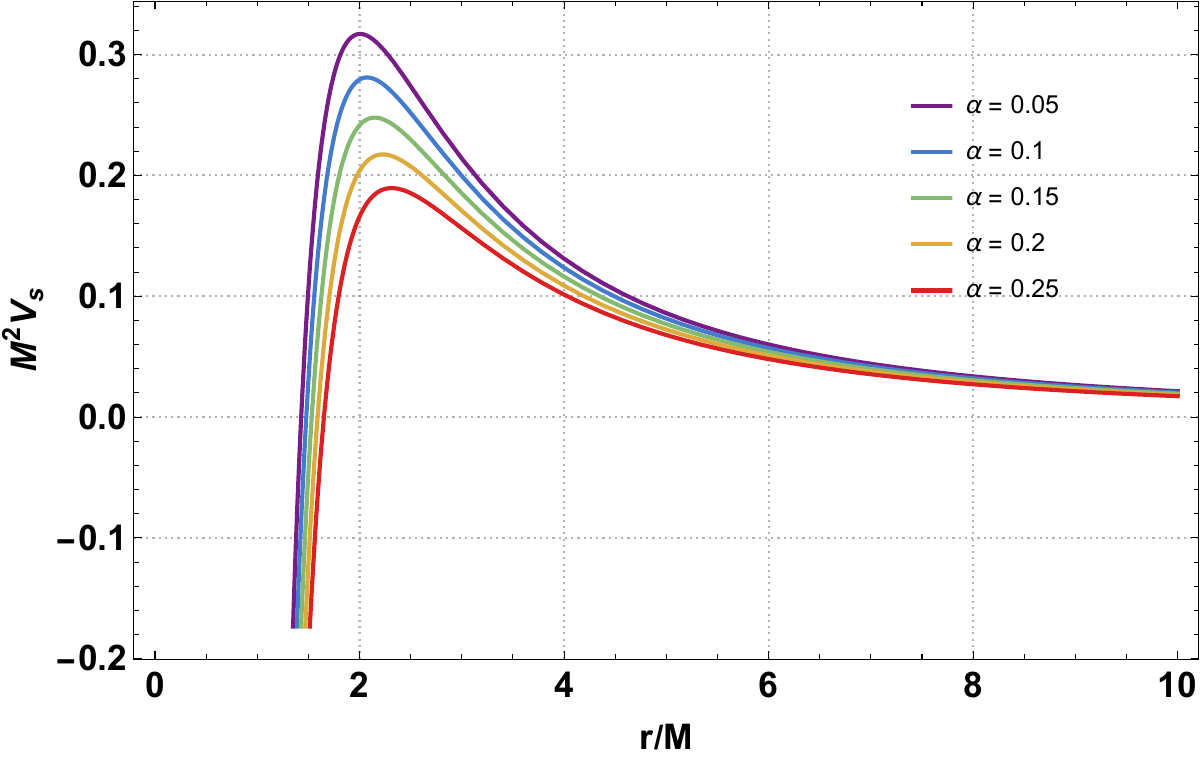}\\
    (a) $\alpha=0.1$ \hspace{6cm} (b) $\beta/M=1.5$
    \caption{Scalar perturbation potential $M^2 V_{\rm s}$ as a function of $r/M$. Panel (a): varying $\beta/M = 1.0, 1.2, 1.4, 1.6, 1.8$ with fixed $\alpha = 0.1$. Panel (b): varying $\alpha = 0.05, 0.1, 0.15, 0.2, 0.25$ with fixed $\beta/M = 1.5$. Other parameters: $Q/M=1$, $q/M=0.1$, $\ell=1$. Both $\beta/M$ and $\alpha$ lower the potential peak, reducing the gravitational barrier for scalar waves.}
    \label{fig:scalar}
\end{figure}

Figure~\ref{fig:scalar} displays the scalar perturbation potential as a function of $r/M$ for various parameter combinations. Increasing $\beta/M$ and $\alpha$ lowers the potential peak height, indicating a reduced gravitational barrier for the massless scalar field. This suppression suggests that scalar waves propagate more easily through the BH spacetime for larger values of these parameters, potentially leading to longer-lived QNMs and slower decay of scalar perturbations.

The QNMs are defined by imposing purely ingoing waves at the horizon and purely outgoing waves at infinity:
\begin{equation}
\psi \sim \begin{cases}
e^{-i\omega r_*} & \text{as } r_* \to -\infty \text{ (horizon)}, \\
e^{+i\omega r_*} & \text{as } r_* \to +\infty \text{ (infinity)}.
\end{cases}
\label{ss_bc}
\end{equation}
These boundary conditions select a discrete set of complex frequencies $\omega_n = \omega_R + i\omega_I$, where $\omega_R$ represents the oscillation frequency and $\omega_I < 0$ (for stable modes) gives the damping rate. The QNM spectrum can be computed using various methods including the WKB approximation, continued fraction method, or time-domain integration \cite{isz38}.

The greybody factor quantifies the probability that a wave escaping from near the horizon reaches infinity, accounting for backscattering by the effective potential barrier. We investigate the transmission and reflection probabilities using a semi-analytical approach that provides rigorous bounds for one-dimensional potential scattering \cite{isz42,isz43,gf4}.

Following Refs.~\cite{isz42,isz43,gf4,gf5,gf6}, the lower bound on the transmission probability (greybody factor) is
\begin{equation}
T(\omega) \geq \mathrm{sech}^2\left(\int_{-\infty}^{+\infty} \wp\,dr_*\right),\label{kk1}
\end{equation}
and the upper bound on the reflection probability is
\begin{equation}
R(\omega) \leq \tanh^2\left(\int_{-\infty}^{+\infty} \wp\,dr_*\right),\label{kk2}
\end{equation}
where the function $\wp$ is defined as
\begin{equation}
\wp = \frac{\sqrt{(H')^2 + (\omega^2 - V_{\rm eff} - H^2)^2}}{2H}.\label{kk3}
\end{equation}
Here, $H$ is a positive function satisfying $H(r_*) > 0$ and $H(\pm\infty) = \omega$, and $V_{\rm eff}$ denotes the effective potential.

Choosing $H^2 = \omega^2 - V_{\rm eff}$ in Eq.~\eqref{kk3}, the bounds simplify to
\begin{equation}
T(\omega) \geq \mathrm{sech}^2\left(\frac{1}{2}\int_{-\infty}^{+\infty}\left|\frac{H'}{H}\right|dr_*\right),\label{kk4}
\end{equation}
\begin{equation}
R(\omega) \leq \tanh^2\left(\frac{1}{2}\int_{-\infty}^{+\infty}\left|\frac{H'}{H}\right|dr_*\right).\label{kk5}
\end{equation}

To avoid divergences, the integration domain is divided into three regions around the potential extrema. The transmission and reflection probabilities can then be expressed in terms of the peak potential value $V_{\rm peak}$ as \cite{gf5}:
\begin{equation}
T(\omega) \geq \frac{4\omega^2(\omega^2 - V_{\rm peak})}{(2\omega^2 - V_{\rm peak})^2},\label{kk6}
\end{equation}
\begin{equation}
R(\omega) \leq \frac{V_{\rm peak}^2}{(2\omega^2 - V_{\rm peak})^2}.\label{kk7}
\end{equation}

\begin{figure}[ht!]
    \centering
    \includegraphics[width=0.45\linewidth]{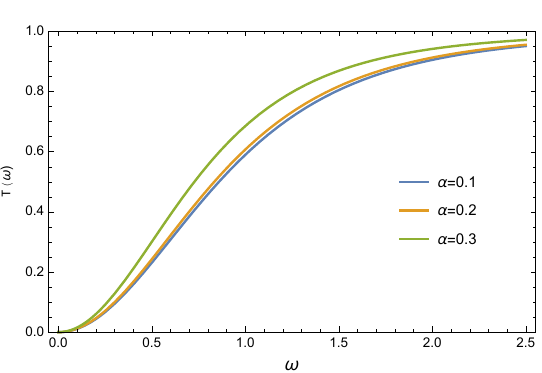}\quad
    \includegraphics[width=0.45\linewidth]{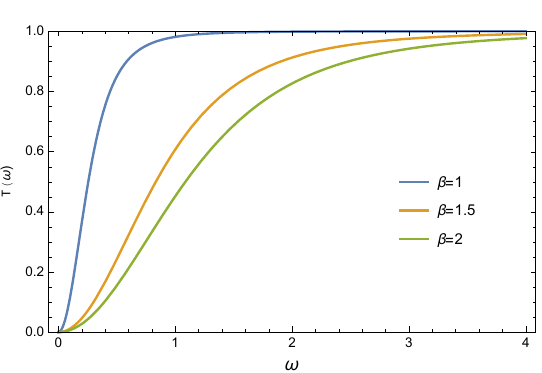}\\
    (a) $\beta/M=1$ \hspace{6cm} (b) $\alpha=0.2$
    \caption{Transmission probability $T(\omega)$ of the charged Bardeen BH with PFDM and CS. Panel (a): varying $\alpha = 0.1, 0.2, 0.3$ with fixed $\beta/M = 1$. Panel (b): varying $\beta = 1, 1.5, 2$ with fixed $\alpha = 0.2$. Other parameters: $M=1$, $Q=1$, $q=0.1$. Increasing $\alpha$ enhances transmission, while increasing $\beta$ suppresses it.}
    \label{figA01}
\end{figure}

\begin{figure}[ht!]
    \centering
    \includegraphics[width=0.45\linewidth]{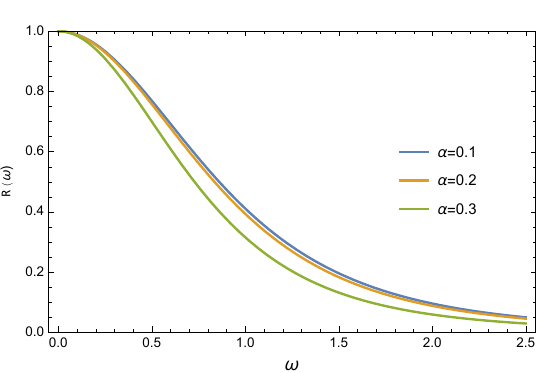}\quad
    \includegraphics[width=0.45\linewidth]{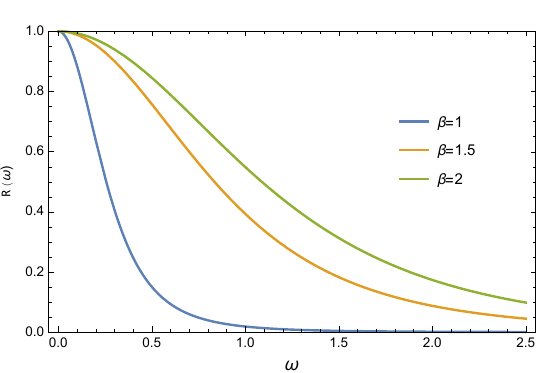}\\
    (a) $\beta/M=1$ \hspace{6cm} (b) $\alpha=0.2$
    \caption{Reflection probability $R(\omega)$ of the charged Bardeen BH with PFDM and CS. Panel (a): varying $\alpha = 0.1, 0.2, 0.3$ with fixed $\beta/M = 1$. Panel (b): varying $\beta = 1, 1.5, 2$ with fixed $\alpha = 0.2$. Other parameters: $M=1$, $Q=1$, $q=0.1$. The reflection behavior is complementary to transmission: increasing $\alpha$ suppresses reflection, while increasing $\beta$ enhances it.}
    \label{figA02}
\end{figure}

Figures~\ref{figA01} and \ref{figA02} display the transmission and reflection probabilities as functions of the frequency $\omega$. The analysis reveals distinct effects of the CS and PFDM parameters:

\textbf{(i) CS parameter effect:} Increasing $\alpha$ enhances the transmission probability and suppresses reflection. This behavior correlates with the reduced potential peak height observed in Fig.~\ref{fig:scalar}(b), which lowers the effective barrier for scalar wave propagation.

\textbf{(ii) PFDM parameter effect:} Increasing $\beta$ suppresses transmission and enhances reflection. Despite both parameters lowering the potential peak, PFDM modifies the potential shape differently, particularly through the logarithmic term that affects the asymptotic behavior.

Table~\ref{tab:scalar_summary} summarizes the influence of the CS and PFDM parameters on scalar perturbation characteristics. The table reveals a notable dichotomy in how these two parameters affect wave propagation through the BH spacetime. While both $\alpha$ and $\beta$ lower the effective potential peak $V_{\rm peak}$, their effects on transmission and reflection probabilities are opposite: increasing the CS parameter $\alpha$ facilitates wave escape (higher $T$, lower $R$), whereas increasing the PFDM parameter $\beta$ impedes it (lower $T$, higher $R$). This counterintuitive behavior arises from the distinct ways these parameters modify the potential shape. The CS contribution appears as a constant shift $(1-\alpha)$ that uniformly lowers the barrier without significantly altering its width, thereby reducing backscattering. In contrast, the PFDM logarithmic term $(\beta/r)\ln(r/\beta)$ modifies the potential profile at intermediate and large radii, effectively broadening the barrier region and enhancing wave reflection despite the reduced peak height. The expected decrease in the QNM damping rate $|\omega_I|$ for both parameters suggests that scalar perturbations decay more slowly in spacetimes with stronger CS or PFDM contributions, potentially leaving longer ringdown signatures in GW observations.

\begin{table}[ht!]
\centering
\setlength{\tabcolsep}{10pt}
\renewcommand{\arraystretch}{1.5}
\begin{tabular}{|l|c|c|p{5.5cm}|}
\hline
\rowcolor{orange!50}
\textbf{Property} & \textbf{Effect of $\alpha \uparrow$} & \textbf{Effect of $\beta \uparrow$} & \textbf{Physical Implication} \\
\hline
Potential peak $V_{\rm peak}$ & Decreases & Decreases & Lower gravitational barrier; waves encounter reduced resistance near the PS \\
\hline
Transmission $T(\omega)$ & Increases & Decreases & CS facilitates wave escape; PFDM impedes propagation to infinity \\
\hline
Reflection $R(\omega)$ & Decreases & Increases & Less backscattering with CS; enhanced reflection in PFDM environment \\
\hline
QNM damping rate $|\omega_I|$ & Expected decrease & Expected decrease & Longer-lived oscillation modes; extended ringdown phase in GW signals \\
\hline
\end{tabular}
\caption{Effects of CS parameter $\alpha$ and PFDM parameter $\beta$ on scalar perturbation properties. Despite both parameters lowering the potential peak, they produce opposite effects on transmission and reflection due to their distinct modifications of the potential profile: CS provides a uniform shift while PFDM introduces logarithmic corrections that broaden the effective barrier.}
\label{tab:scalar_summary}
\end{table}

The observations in Table~\ref{tab:scalar_summary} carry important implications for distinguishing BHs surrounded by different matter environments. In the context of GW astronomy, the ringdown phase following BH mergers encodes the QNM spectrum. The opposite responses of $T(\omega)$ and $R(\omega)$ to $\alpha$ and $\beta$ suggest that combined measurements of greybody factors (through Hawking radiation spectra, if detectable) and QNM frequencies could in principle allow independent determination of the CS and PFDM parameters. Furthermore, the longer-lived modes predicted for both parameters imply that the late-time tail of the ringdown signal may be enhanced relative to vacuum BH solutions. The distinct responses of transmission and reflection to $\alpha$ and $\beta$ might provide potential observational signatures \cite{isz39,isz40}.

\section{Conclusion} \label{isec7}

In this work, we investigated the gravitational and dynamical properties of a charged Bardeen BH surrounded by PFDM and coupled to a CS. The spacetime geometry, described by the metric function in Eq.~\eqref{function}, incorporates five parameters: the BH mass $M$, the magnetic monopole charge $q$ arising from NED, the electric charge $Q$, the CS parameter $\alpha$, and the PFDM parameter $\beta$. This multi-parameter configuration allowed us to explore how exotic matter distributions and topological defects modify the classical BH solutions and their observable signatures.

In Sec.~\ref{isec2}, we presented the theoretical framework combining GR with NED, the Nambu-Goto action for the CS, and the anisotropic energy-momentum tensor characterizing PFDM. The resulting metric function exhibits several notable features: the Bardeen-type magnetic charge regularizes the geometry near $r = 0$, preventing the formation of a curvature singularity; the PFDM introduces a logarithmic correction $(\beta/r)\ln(r/\beta)$ that dominates at intermediate and large radii; and the CS contribution shifts the asymptotic value of the metric to $f(r \to \infty) = 1 - \alpha$, rendering the spacetime asymptotically non-flat. We analyzed the horizon structure by numerically solving $f(r_h) = 0$ for various parameter combinations. The results, summarized in Table~\ref{tab:horizons} and visualized in Fig.~\ref{fig:metric_function}, revealed four distinct causal configurations: non-extremal BHs with two horizons ($r_- < r_+$), extremal BHs where the horizons coincide, single-horizon BHs analogous to Schwarzschild, and naked singularities where no horizon exists. We found that the CS parameter $\alpha$ shifts the outer horizon outward while compressing the inner horizon, whereas the PFDM parameter $\beta$ widens the separation between the two horizons. Large values of the charge parameters $q$ and $Q$ can eliminate horizons entirely, producing naked singularity configurations.

In Sec.~\ref{isec3}, we studied null geodesics and the PS properties using the Lagrangian formalism. The effective potential for photons, given in Eq.~\eqref{bb7}, determines the gravitational barrier that light must overcome to escape from the BH vicinity. We showed in Fig.~\ref{fig:null} that both $\alpha$ and $\beta$ suppress the potential peak, reducing the barrier height and allowing photons with lower impact parameters to escape. The PS radius $r_{\rm ph}$, obtained from the condition $\partial V_{\rm eff}/\partial r = 0$ in Eq.~\eqref{bb12}, increases monotonically with both $\alpha$ and $\beta$, as demonstrated in Tables~\ref{tab:1}--\ref{tab:2} and Fig.~\ref{fig:1}. For the BH shadow observable by distant observers, we employed the formula appropriate for asymptotically non-flat spacetimes, Eq.~\eqref{bb16}, which includes the factor $(1-\alpha)^{1/2}$ accounting for the CS modification. The shadow radius $R_{\rm sh}$ also increases with both parameters (Table~\ref{tab:shadow2} and Fig.~\ref{fig:shadow2}), suggesting that BHs embedded in PFDM and CS environments cast larger shadows compared to their vacuum counterparts. This result has potential implications for EHT observations of supermassive BHs, where deviations from the Kerr shadow could signal the presence of surrounding dark matter or string cloud configurations \cite{isz03,isz03b,isz03c,isz04,isz04b,isz04c}.

Section~\ref{isec4} addressed the dynamics of neutral test particles through the Hamiltonian formalism. The effective potential for timelike geodesics, Eq.~\eqref{cc10}, exhibits different behavior compared to the null case: increasing $\beta$ raises the potential (enhancing the barrier), while increasing $\alpha$ lowers it. This opposite response, illustrated in Fig.~\ref{fig:timelike}, indicates that the PFDM and CS affect massive and massless particles differently. We derived the specific angular momentum $\mathcal{L}_{\rm sp}$ and specific energy $\mathcal{E}_{\rm sp}$ for circular orbits in Eqs.~\eqref{cc13}--\eqref{cc14}. The numerical analysis showed that both parameters increase the angular momentum required for circular orbits (Fig.~\ref{fig:specific-momentum}), while their effects on the specific energy differ: $\beta$ increases $\mathcal{E}_{\rm sp}$, whereas $\alpha$ decreases it at fixed $\beta$ (Fig.~\ref{fig:specific-energy}). The ISCO location, determined by Eq.~\eqref{cc26}, shifts outward with increasing $\alpha$ and $\beta$, affecting the inner edge of accretion disks and consequently the thermal spectrum observed in X-ray binaries.

In Sec.~\ref{isec5}, we examined QPOs, which serve as probes of strong-field gravity near compact objects. We computed the azimuthal frequency $\Omega_\phi$ in Eq.~\eqref{pp2} and found that it depends on $\beta$, $q$, and $Q$, but is independent of the CS parameter $\alpha$. This independence provides a potential method to disentangle the effects of PFDM and CS through multi-frequency QPO observations. The radial and vertical epicyclic frequencies, Eqs.~\eqref{pp3}--\eqref{pp4}, respond differently to the spacetime parameters: both $\alpha$ and $\beta$ increase $\Omega_r$ (Fig.~\ref{fig:radial-frequency}) but decrease $\Omega_\theta$ (Fig.~\ref{fig:vertical-frequency}). This asymmetry could produce characteristic QPO frequency ratios distinguishable from standard BH models. The periastron precession frequency $\Omega_p = \Omega_\phi - \Omega_r$, plotted in Fig.~\ref{fig:periastron-frequency}, decreases with $\beta$, indicating slower orbital precession in PFDM-rich environments. These frequency modifications may be detectable in high-precision timing observations of accreting stellar-mass BHs \cite{isz04}.

Section~\ref{isec6} focused on scalar perturbations and their QNM spectrum. The Klein-Gordon equation on the curved background reduces to a Schrödinger-like equation with an effective potential $V_{\rm s}(r)$ given in Eq.~\eqref{ss4}. Similar to the null geodesic case, both $\alpha$ and $\beta$ lower the potential peak (Fig.~\ref{fig:scalar}), suggesting that scalar waves propagate more easily in spacetimes with stronger PFDM or CS contributions. We analyzed the greybody factors using the semi-analytical bounds of Eqs.~\eqref{kk6}--\eqref{kk7}. The transmission and reflection probabilities, displayed in Figs.~\ref{figA01}--\ref{figA02}, revealed an interesting dichotomy: increasing $\alpha$ enhances transmission (reduces reflection), while increasing $\beta$ suppresses transmission (enhances reflection). This opposite behavior, despite both parameters lowering the potential peak, arises from the different ways the CS and PFDM modify the potential shape, particularly through the logarithmic term affecting the asymptotic region. The QNM spectrum, characterized by complex frequencies $\omega = \omega_R + i\omega_I$, encodes information about the BH ringdown following perturbations. The lower potential barriers suggest longer-lived modes (smaller $|\omega_I|$), which could leave imprints in the GW signals detected by LIGO-Virgo-KAGRA during BH merger events \cite{isz05,isz06}.

Throughout our analysis, we identified several observational signatures that could distinguish charged Bardeen BHs with PFDM and CS from standard GR solutions. The enlarged shadow radius, modified QPO frequency ratios, shifted ISCO location, and altered greybody factors all provide potential tests of these exotic BH configurations. The fact that $\alpha$ and $\beta$ produce different (and sometimes opposite) effects on various observables offers the possibility of independently constraining these parameters through combined multi-messenger observations.

Looking ahead, several extensions of this work merit further investigation. First, the inclusion of BH rotation through a Kerr-like generalization of the metric would allow comparison with EHT observations of M87* and Sgr A*, both of which exhibit non-zero spin. Second, the computation of QNM frequencies using the WKB method or time-domain integration would provide quantitative predictions for GW ringdown signals. Third, the study of electromagnetic perturbations and their coupling to the NED sector could reveal additional signatures of the magnetic monopole charge. Fourth, constraints on the parameter space using current EHT shadow measurements and X-ray timing data from NICER and RXTE would establish observational bounds on $\alpha$, $\beta$, $q$, and $Q$. Finally, the thermodynamic properties of the charged Bardeen BH with PFDM and CS, including Hawking temperature, entropy, and phase transitions, remain to be explored in detail. These directions will help bridge the gap between theoretical BH models incorporating dark matter and string theory-inspired corrections and the growing wealth of observational data from electromagnetic and GW astronomy.
}

\small

\section*{Acknowledgments}
F.A. gratefully acknowledges the Inter University Centre for Astronomy and Astrophysics (IUCAA), Pune, India, for the award of a visiting associateship. \.{I}.~S. is thankful for academic support provided by EMU, T\"{U}B\.{I}TAK, ANKOS, and SCOAP3, as well as for networking support received through COST Actions CA22113, CA21106, CA23130, and CA23115.

\bibliographystyle{unsrtnat} 
\bibliography{ref}

\begin{thebibliography}{93}
\providecommand{\natexlab}[1]{#1}
\providecommand{\url}[1]{\texttt{#1}}
\expandafter\ifx\csname urlstyle\endcsname\relax
  \providecommand{\doi}[1]{doi: #1}\else
  \providecommand{\doi}{doi: \begingroup \urlstyle{rm}\Url}\fi

\bibitem[Abbott et~al.(2016)]{isz01}
B.~P. Abbott et~al.
\newblock Observation of gravitational waves from a binary black hole merger.
\newblock \emph{Phys. Rev. Lett.}, 116:\penalty0 061102, 2016.
\newblock \doi{10.1103/PhysRevLett.116.061102}.

\bibitem[Abbott et~al.(2023)]{isz02}
R.~Abbott et~al.
\newblock Gwtc-3: Compact binary coalescences observed by ligo and virgo during
  the second part of the third observing run.
\newblock \emph{Phys. Rev. X}, 13:\penalty0 041039, 2023.
\newblock \doi{10.1103/PhysRevX.13.041039}.

\bibitem[Akiyama et~al.(2019{\natexlab{a}})]{isz03}
K.~Akiyama et~al.
\newblock First m87 event horizon telescope results. i. the shadow of the
  supermassive black hole.
\newblock \emph{Astrophys. J. Lett.}, 875:\penalty0 L1, 2019{\natexlab{a}}.
\newblock \doi{10.3847/2041-8213/ab0ec7}.

\bibitem[Akiyama et~al.(2019{\natexlab{b}})]{isz03b}
K.~Akiyama et~al.
\newblock First m87 event horizon telescope results. i. the shadow of the
  supermassive black hole.
\newblock \emph{Astrophys. J. Lett.}, 875:\penalty0 L4, 2019{\natexlab{b}}.
\newblock \doi{10.3847/2041-8213/ab0e85}.

\bibitem[Akiyama et~al.(2019{\natexlab{c}})]{isz03c}
K.~Akiyama et~al.
\newblock First m87 event horizon telescope results. i. the shadow of the
  supermassive black hole.
\newblock \emph{Astrophys. J. Lett.}, 875:\penalty0 L6, 2019{\natexlab{c}}.
\newblock \doi{10.3847/2041-8213/ab1141}.

\bibitem[Akiyama et~al.(2022{\natexlab{a}})]{isz04}
K.~Akiyama et~al.
\newblock First sagittarius a* event horizon telescope results. i. the shadow
  of the supermassive black hole.
\newblock \emph{Astrophys. J. Lett.}, 930:\penalty0 L12, 2022{\natexlab{a}}.
\newblock \doi{10.3847/2041-8213/ac6674}.

\bibitem[Akiyama et~al.(2022{\natexlab{b}})]{isz04b}
K.~Akiyama et~al.
\newblock First sagittarius a* event horizon telescope results. i. the shadow
  of the supermassive black hole.
\newblock \emph{Astrophys. J. Lett.}, 930:\penalty0 L14, 2022{\natexlab{b}}.
\newblock \doi{10.3847/2041-8213/ac6429}.

\bibitem[Akiyama et~al.(2022{\natexlab{c}})]{isz04c}
K.~Akiyama et~al.
\newblock First sagittarius a* event horizon telescope results. i. the shadow
  of the supermassive black hole.
\newblock \emph{Astrophys. J. Lett.}, 930:\penalty0 L17, 2022{\natexlab{c}}.
\newblock \doi{10.3847/2041-8213/ac6756}.

\bibitem[Hawking and Penrose(1970)]{isz05}
S.~W. Hawking and R.~Penrose.
\newblock The singularities of gravitational collapse and cosmology.
\newblock \emph{Proc. R. Soc. Lond. A}, 314:\penalty0 529--548, 1970.
\newblock \doi{10.1098/rspa.1970.0021}.

\bibitem[Penrose(1965)]{isz06}
R.~Penrose.
\newblock Gravitational collapse and space-time singularities.
\newblock \emph{Phys. Rev. Lett.}, 14:\penalty0 57--59, 1965.
\newblock \doi{10.1103/PhysRevLett.14.57}.

\bibitem[Bardeen(1968)]{isz07}
J.~Bardeen.
\newblock {Non-singular general relativistic gravitational collapse}.
\newblock In \emph{Proceedings of the 5th International Conference on
  Gravitation and the Theory of Relativity}, page~87, September 1968.

\bibitem[Ay{\'{o}}n-Beato and Garc{\'{\i}}a(2000)]{isz08}
E.~Ay{\'{o}}n-Beato and A.~Garc{\'{\i}}a.
\newblock The bardeen model as a nonlinear magnetic monopole.
\newblock \emph{Phys. Lett. B}, 493:\penalty0 149--152, 2000.
\newblock \doi{10.1016/S0370-2693(00)01125-4}.

\bibitem[Ay{\'{o}}n-Beato and Garc{\'{\i}}a(1998)]{isz09}
E.~Ay{\'{o}}n-Beato and A.~Garc{\'{\i}}a.
\newblock Regular black hole in general relativity coupled to nonlinear
  electrodynamics.
\newblock \emph{Phys. Rev. Lett.}, 80:\penalty0 5056--5059, 1998.
\newblock \doi{10.1103/PhysRevLett.80.5056}.

\bibitem[Ansoldi(2008)]{isz10}
S.~Ansoldi.
\newblock Spherical black holes with regular center: A review of existing
  models including a recent realization with gaussian sources, 2008.

\bibitem[Bronnikov()]{isz11}
K.~A. Bronnikov.
\newblock Regular magnetic black holes and monopoles from nonlinear
  electrodynamics.
\newblock \emph{Phys. Rev. D}.
\newblock \doi{10.1103/PhysRevD.63.044005}.

\bibitem[Bambi and Modesto(2013)]{isz12}
C.~Bambi and L.~Modesto.
\newblock Rotating regular black holes.
\newblock \emph{Phys. Lett. B}, 721:\penalty0 329--334, 2013.
\newblock \doi{10.1016/j.physletb.2013.03.025}.

\bibitem[Rodrigues et~al.(2016)Rodrigues, Junior, Marques, and Zanchin]{isz13}
M.~E. Rodrigues, E.~L.~B. Junior, G.~T. Marques, and V.~T. Zanchin.
\newblock Regular black holes in {f(R)} gravity coupled to nonlinear
  electrodynamics.
\newblock \emph{Phys. Rev. D}, 94:\penalty0 024062, 2016.
\newblock \doi{10.1103/PhysRevD.94.024062}.

\bibitem[Toshmatov et~al.(2017)Toshmatov, Stuchl{\'{\i}}k, and Ahmedov]{isz14}
B.~Toshmatov, Z.~Stuchl{\'{\i}}k, and B.~Ahmedov.
\newblock Generic rotating regular black holes in general relativity coupled to
  nonlinear electrodynamics.
\newblock \emph{Phys. Rev. D}, 95:\penalty0 084037, 2017.
\newblock \doi{10.1103/PhysRevD.95.084037}.

\bibitem[Fernando and Correa(2012)]{isz15}
S.~Fernando and J.~Correa.
\newblock Quasinormal modes of bardeen black hole: Scalar perturbations.
\newblock \emph{Phys. Rev. D}, 86:\penalty0 064039, 2012.
\newblock \doi{10.1103/PhysRevD.86.064039}.

\bibitem[Ahmed et~al.(2025{\natexlab{a}})Ahmed, Al-Badawi, İzzet Sakallı, and
  Kanzi]{AHMED2025101907}
Faizuddin Ahmed, Ahmad Al-Badawi, İzzet Sakallı, and Sara Kanzi.
\newblock Motions of test particles in gravitational field, perturbations and
  greybody factor of bardeen-like ads black hole with phantom global monopoles.
\newblock \emph{Phys. Dark Univ.}, 48:\penalty0 101907, 2025{\natexlab{a}}.
\newblock \doi{10.1016/j.dark.2025.101907}.

\bibitem[Al-Badawi et~al.(2020)Al-Badawi, İzzet Sakallı, and
  Kanzi]{ALBADAWI2020168026}
Ahmad Al-Badawi, İzzet Sakallı, and Sara Kanzi.
\newblock Solution of dirac equation and greybody radiation around a regular
  bardeen black hole surrounded by quintessence.
\newblock \emph{Ann. Phys. (NY)}, 412:\penalty0 168026, 2020.
\newblock \doi{10.1016/j.aop.2019.168026}.

\bibitem[Sucu et~al.(2025)Sucu, Sakall{\i}, Sert, and Sucu]{Sucu:2025fwa}
Erdem Sucu, {\.I}zzet Sakall{\i}, {\"O}zcan Sert, and Yusuf Sucu.
\newblock {Quantum-corrected thermodynamics and plasma lensing in non-minimally
  coupled symmetric teleparallel black holes}.
\newblock \emph{Phys. Dark Univ.}, 50:\penalty0 102063, 2025.
\newblock \doi{10.1016/j.dark.2025.102063}.

\bibitem[Aydiner et~al.(2025)Aydiner, Sucu, and Sakall{\i}]{Aydiner:2025eii}
Ekrem Aydiner, Erdem Sucu, and {\.I}zzet Sakall{\i}.
\newblock {Regular magnetically charged black holes from nonlinear
  electrodynamics: Thermodynamics, light deflection, and orbital dynamics}.
\newblock \emph{Phys. Dark Univ.}, 50:\penalty0 102164, 2025.
\newblock \doi{10.1016/j.dark.2025.102164}.

\bibitem[G{\"u}rsel et~al.(2025)G{\"u}rsel, Mangut, and Sucu]{Gursel:2025wan}
Huriye G{\"u}rsel, Mert Mangut, and Erdem Sucu.
\newblock {Thermodynamics of Einstein-Euler-Heisenberg Black Holes with Thermal
  Fluctuations and Nonlinear Electromagnetic Fields}.
\newblock \emph{Class. Quant. Grav.}, 42:\penalty0 135015, 2025.
\newblock \doi{10.1088/1361-6382/ade7ea}.

\bibitem[Sucu(2026)]{Sucu:2025pce}
Erdem Sucu.
\newblock {Quantum gravity corrections and plasma-induced lensing of
  magnetically charged black holes}.
\newblock \emph{Nucl. Phys. B}, 1022:\penalty0 117285, 2026.
\newblock \doi{10.1016/j.nuclphysb.2025.117285}.

\bibitem[Sakall{\i} et~al.(2026)Sakall{\i}, Sucu, and Sucu]{Sakalli:2025els}
{\.I}zzet Sakall{\i}, Yusuf Sucu, and Erdem Sucu.
\newblock {Zitterbewegung oscillations and GUP-induced quantum modifications of
  Yang-Mills black holes in perfect fluid dark matter}.
\newblock \emph{Nucl. Phys. B}, 1022:\penalty0 117216, 2026.
\newblock \doi{10.1016/j.nuclphysb.2025.117216}.

\bibitem[Sucu and Sakall{\i}(2025)]{Sucu:2025eix}
Erdem Sucu and {\.I}zzet Sakall{\i}.
\newblock {Charged regular black holes in quantum gravity: from thermodynamic
  stability to observational phenomena}.
\newblock \emph{Eur. Phys. J. C}, 85\penalty0 (9):\penalty0 989, 2025.
\newblock \doi{10.1140/epjc/s10052-025-14726-5}.

\bibitem[Ade et~al.(2016)]{isz16}
P.~A.~R. Ade et~al.
\newblock Planck 2015 results. xiii. cosmological parameters.
\newblock \emph{Astron. Astrophys.}, 594:\penalty0 A13, 2016.
\newblock \doi{10.1051/0004-6361/201525830}.

\bibitem[Bertone and Hooper(2018)]{isz17}
G.~Bertone and D.~Hooper.
\newblock History of dark matter.
\newblock \emph{Rev. Mod. Phys.}, 90:\penalty0 045002, 2018.
\newblock \doi{10.1103/RevModPhys.90.045002}.

\bibitem[Kiselev(2003)]{isz18}
V.~V. Kiselev.
\newblock Quintessence and black holes.
\newblock \emph{Class. Quantum Grav.}, 20:\penalty0 1187--1198, 2003.
\newblock \doi{10.1088/0264-9381/20/6/310}.

\bibitem[Li and Yang(2012)]{isz19}
M.-H. Li and K.-C. Yang.
\newblock Galactic dark matter in the phantom field.
\newblock \emph{Phys. Rev. D}, 86:\penalty0 123015, 2012.
\newblock \doi{10.1103/PhysRevD.86.123015}.

\bibitem[Xu et~al.(2018{\natexlab{a}})Xu, Hou, Gong, and Wang]{isz20}
Z.~Xu, X.~Hou, X.~Gong, and J.~Wang.
\newblock Kerr--newman--ads black hole surrounded by perfect fluid matter in
  rastall gravity.
\newblock \emph{Eur. Phys. J. C}, 78:\penalty0 513, 2018{\natexlab{a}}.
\newblock \doi{10.1140/epjc/s10052-018-5991-x}.

\bibitem[Ghosh et~al.(2018)Ghosh, Singh, and Maharaj]{isz21}
S.~G. Ghosh, D.~V. Singh, and S.~D. Maharaj.
\newblock Regular black holes in einstein--gauss--bonnet gravity.
\newblock \emph{Phys. Rev. D}, 97:\penalty0 104050, 2018.
\newblock \doi{10.1103/PhysRevD.97.104050}.

\bibitem[Zhang et~al.(2021)Zhang, Chen, Ma, He, and Deng]{isz22}
H.-X. Zhang, Y.~Chen, T.-C. Ma, P.-Z. He, and J.-B. Deng.
\newblock Bardeen black hole surrounded by perfect fluid dark matter.
\newblock \emph{Chin. Phys. C}, 45:\penalty0 055103, 2021.
\newblock \doi{10.1088/1674-1137/abe7f5}.

\bibitem[Haroon et~al.(2019)Haroon, Jamil, Jusufi, Lin, and Mann]{isz23}
S.~Haroon, M.~Jamil, K.~Jusufi, K.~Lin, and R.~B. Mann.
\newblock Shadow and deflection angle of rotating black holes in perfect fluid
  dark matter with a cosmological constant.
\newblock \emph{Phys. Rev. D}, 99:\penalty0 044015, 2019.
\newblock \doi{10.1103/PhysRevD.99.044015}.

\bibitem[Xu et~al.(2018{\natexlab{b}})Xu, Hou, and Wang]{Xu2018_Kerr_PFDM}
Zhaoyi Xu, Xian Hou, and Jiancheng Wang.
\newblock Kerr--anti-de sitter/de sitter black hole in perfect fluid dark
  matter background.
\newblock \emph{Class. Quantum Grav.}, 35:\penalty0 115003, 2018{\natexlab{b}}.
\newblock \doi{10.1088/1361-6382/aabcb6}.

\bibitem[Rizwan et~al.(2019)Rizwan, Jamil, and Jusufi]{Rizwan2019_PFDM_Kerr}
Muhammad Rizwan, Mubasher Jamil, and Kimet Jusufi.
\newblock Distinguishing a kerr-like black hole and a naked singularity in
  perfect fluid dark matter via precession frequencies.
\newblock \emph{Phys. Rev. D}, 99:\penalty0 024050, 2019.
\newblock \doi{10.1103/PhysRevD.99.024050}.

\bibitem[Hendi et~al.(2020)Hendi, Nemati, Lin, and Jamil]{Hendi2020_PFDM}
Seyed~Hossein Hendi, Azadeh Nemati, Kai Lin, and Mubasher Jamil.
\newblock Instability and phase transitions of a rotating black hole in the
  presence of perfect fluid dark matter.
\newblock \emph{Eur. Phys. J. C}, 80:\penalty0 296, 2020.
\newblock \doi{10.1140/epjc/s10052-020-7829-6}.

\bibitem[Rizwan and Jusufi(2023)]{Rizwan2023_PFDM}
Muhammad Rizwan and Kimet Jusufi.
\newblock Topological classes of thermodynamics of black holes in perfect fluid
  dark matter background.
\newblock \emph{Eur. Phys. J. C}, 83:\penalty0 944, 2023.
\newblock \doi{10052-023-12126-1}.

\bibitem[Liang et~al.(2023)Liang, Hu, Wu, and An]{Liang2023_PFDM}
Xiao Liang, Ya-Peng Hu, Chen-Hao Wu, and Yu-Sen An.
\newblock Thermodynamics and evaporation of perfect fluid dark matter black
  hole in phantom background.
\newblock \emph{Eur. Phys. J. C}, 83:\penalty0 1009, 2023.
\newblock \doi{10.1140/epjc/s10052-023-12200-8}.

\bibitem[Rizwan et~al.(2025)Rizwan, Jamil, and Moughal]{Rizwan2025_PFDM}
Muhammad Rizwan, Mubasher Jamil, and M.~Z.~A. Moughal.
\newblock Universal thermodynamic topological classes of black holes in a
  perfect fluid dark matter background.
\newblock \emph{Eur. Phys. J. C}, 85:\penalty0 359, 2025.
\newblock \doi{10.1140/epjc/s10052-025-14070-8}.

\bibitem[Sekhmani et~al.(2025)Sekhmani, Maurya, Jasim, Sakall{\i}, Rayimbaev,
  and Ibragimov]{Sekhmani2025_PFDM_ModMax}
Y.~Sekhmani, S.~K. Maurya, M.~K. Jasim, {\.I}.~Sakall{\i}, J.~Rayimbaev, and
  I.~Ibragimov.
\newblock Thermodynamics and phase transition of anti-de sitter black holes
  with modmax nonlinear electrodynamics and perfect fluid dark matter.
\newblock \emph{Eur. Phys. J. C}, 85:\penalty0 229, 2025.
\newblock \doi{10.1140/epjc/s10052-025-13932-5}.

\bibitem[Tan et~al.(2025)Tan, Liu, Liang, and Long]{QT2025}
Q.~Tan, D.~Liu, J.~Liang, and Z.-W. Long.
\newblock Testing black holes in a perfect fluid dark matter environment using
  quasinormal modes.
\newblock \emph{Eur. Phys. J. C}, 85:\penalty0 687, 2025.
\newblock \doi{10.1140/epjc/s10052-025-14407-3}.

\bibitem[Ali et~al.(2025)Ali, Negi, and Pant]{Ali2025_PFDM}
Md~Sabir Ali, Abhishek Negi, and Sanjay Pant.
\newblock Influence of perfect fluid dark matter on shadow observables of
  yang-mills modified charged black holes.
\newblock \emph{arXiv preprint}, 2025.

\bibitem[Ma et~al.(2024)Ma, Wang, Deng, and Hu]{Shi2024_PFDM_EulerHeisenberg}
Shi-Jie Ma, Rui-Bo Wang, Jian-Bo Deng, and Xian-Ru Hu.
\newblock {Euler{\textendash}Heisenberg black hole surrounded by perfect fluid
  dark matter}.
\newblock \emph{Eur. Phys. J. C}, 84\penalty0 (6):\penalty0 595, 2024.
\newblock \doi{10.1140/epjc/s10052-024-12914-3}.

\bibitem[Letelier(1979)]{isz24}
P.~S. Letelier.
\newblock Clouds of strings in general relativity.
\newblock \emph{Phys. Rev. D}, 20:\penalty0 1294--1302, 1979.
\newblock \doi{10.1103/PhysRevD.20.1294}.

\bibitem[Vilenkin and Shellard(1994)]{isz25}
A.~Vilenkin and E.~P.~S. Shellard.
\newblock \emph{Cosmic Strings and Other Topological Defects}.
\newblock Cambridge University Press, Cambridge, 1994.
\newblock ISBN 9780521654760.

\bibitem[Hindmarsh and Kibble(1995)]{isz26}
M.~B. Hindmarsh and T.~W.~B. Kibble.
\newblock Cosmic strings.
\newblock \emph{Rep. Prog. Phys.}, 58:\penalty0 477--562, 1995.
\newblock \doi{10.1088/0034-4885/58/5/001}.

\bibitem[Toledo and Bezerra(2019)]{isz27}
J.~M. Toledo and V.~B. Bezerra.
\newblock Black holes with cloud of strings and a cosmological constant in
  {f(R)} gravity.
\newblock \emph{Eur. Phys. J. C}, 79:\penalty0 110, 2019.
\newblock \doi{10.1140/epjc/s10052-019-6628-4}.

\bibitem[Ganguly et~al.(2014)Ganguly, Ghosh, and Maharaj]{isz28}
A.~Ganguly, S.~G. Ghosh, and S.~D. Maharaj.
\newblock Accretion onto a black hole in a string cloud background.
\newblock \emph{Phys. Rev. D}, 90:\penalty0 064037, 2014.
\newblock \doi{10.1103/PhysRevD.90.064037}.

\bibitem[Rodrigues and Vieira(2022)]{isz29}
M.~E. Rodrigues and H.~A. Vieira.
\newblock Bardeen solution with a cloud of strings.
\newblock \emph{Phys. Rev. D}, 106:\penalty0 084015, 2022.
\newblock \doi{10.1103/PhysRevD.106.084015}.

\bibitem[Sood et~al.(2024)Sood, Ali, Singh, and Ghosh]{isz30}
A.~Sood, M.~S. Ali, J.~K. Singh, and S.~G. Ghosh.
\newblock Photon orbits and phase transition for letelier ads black holes
  immersed in perfect fluid dark matter.
\newblock \emph{Chin. Phys. C}, 48:\penalty0 065109, 2024.
\newblock \doi{10.1088/1674-1137/ad361f}.

\bibitem[Ahmed et~al.(2025{\natexlab{b}})Ahmed, Sakall{\i}, and
  Al-Badawi]{Yang2024_String}
Faizuddin Ahmed, {\.I}zzet Sakall{\i}, and Ahmad Al-Badawi.
\newblock {Kerr-Bertotti-Robinson Black Holes Surrounded by a Cloud of
  Strings}.
\newblock 11 2025{\natexlab{b}}.

\bibitem[do~Nascimento et~al.(2024)do~Nascimento, Bezerra, and
  Toledo]{Nascimento2024_String}
Francinaldo~Florencio do~Nascimento, Valdir~Barbosa Bezerra, and Jefferson
  de~Morais Toledo.
\newblock Black holes with a cloud of strings and quintessence in a non-linear
  electrodynamics scenario.
\newblock \emph{Universe}, 10\penalty0 (11):\penalty0 430, 2024.
\newblock \doi{10.3390/universe10110430}.

\bibitem[Atamurotov et~al.(2023)Atamurotov, Alibekov, Abdujabbarov, Mustafa,
  and Aripov]{Atamurotov2023_String}
Farruh Atamurotov, Husan Alibekov, Ahmadjon Abdujabbarov, Ghulam Mustafa, and
  Mersaid~M. Aripov.
\newblock Weak gravitational lensing around bardeen black hole with a string
  cloud in the presence of plasma.
\newblock \emph{Symmetry}, 15\penalty0 (4):\penalty0 848, 2023.
\newblock \doi{10.3390/sym15040848}.

\bibitem[Ahmed et~al.(2026)Ahmed, Gashti, Bouzenada, and
  Pourhassan]{Ahmed2026_NPB_String}
Faizuddin Ahmed, Saeed~Noori Gashti, Abdelmalek Bouzenada, and Behnam
  Pourhassan.
\newblock Schwarzschild--ads black holes with cloud of strings and
  quintessence: Geodesics and quasinormal modes.
\newblock \emph{Nucl. Phys. B}, 1022:\penalty0 117260, 2026.
\newblock \doi{10.1016/j.nuclphysb.2025.117260}.

\bibitem[Mustafa et~al.(2022)Mustafa, Atamurotov, Hussain, Shaymatov, and
  {\"O}vg{\"u}n]{Mustafa2022_String}
Ghulam Mustafa, Farruh Atamurotov, Ibrar Hussain, Sanjar Shaymatov, and Ali
  {\"O}vg{\"u}n.
\newblock Shadows and gravitational weak lensing by the schwarzschild black
  hole in the string cloud background with quintessential field.
\newblock \emph{Chin. Phys. C}, 46:\penalty0 125107, 2022.
\newblock \doi{10.1088/1674-1137/ac917f}.

\bibitem[Molla et~al.(2024)Molla, Chaudhary, Mustafa, Atamurotov, Debnath, and
  Arora]{Molla2024_String}
Niyaz~Uddin Molla, Himanshu Chaudhary, Ghulam Mustafa, Farruh Atamurotov, Ujjal
  Debnath, and Dhruv Arora.
\newblock Strong gravitational lensing by {Sgr A*} and {M87*} black holes
  embedded in dark matter halo exhibiting string cloud and quintessential
  field.
\newblock \emph{Eur. Phys. J. C}, 84:\penalty0 574, 2024.
\newblock \doi{10.1140/epjc/s10052-024-12504-3}.

\bibitem[Ahmed et~al.(2025{\natexlab{c}})Ahmed, Al-Badawi, and
  Sakall{\i}]{Ahmed2025_String}
Faizuddin Ahmed, Ahmad Al-Badawi, and {\.I}zzet Sakall{\i}.
\newblock Observable signatures of black hole with hernquist dark matter halo
  having a cloud of strings: Geodesic, perturbations, and shadow.
\newblock \emph{Eur. Phys. J. C}, 85:\penalty0 984, 2025{\natexlab{c}}.
\newblock \doi{10.1140/epjc/s10052-025-14266-y}.

\bibitem[Al-Badawi et~al.(2024)Al-Badawi, Shaymatov, Jha, and
  Rahaman]{AlBadawi2024_String}
Ahmad Al-Badawi, Sanjar Shaymatov, Sohan~Kumar Jha, and Anisur Rahaman.
\newblock Gup corrected black holes with cloud of string.
\newblock \emph{Eur. Phys. J. C}, 84:\penalty0 722, 2024.
\newblock \doi{10.1140/epjc/s10052-024-13059-z}.

\bibitem[Waseem et~al.(2023)Waseem, Javed, Gul, Mustafa, and
  Errehymy]{Waseem2023_String}
Arfa Waseem, Faisal Javed, M.~Zeeshan Gul, Ghulam Mustafa, and Abdelghani
  Errehymy.
\newblock Impact of quintessence and cloud of strings on self-consistent
  d-dimensional charged thin-shell wormholes.
\newblock \emph{Eur. Phys. J. C}, 83:\penalty0 1088, 2023.
\newblock \doi{10.1140/epjc/s10052-023-12239-7}.

\bibitem[Nascimento et~al.(2024)Nascimento, Bezerra, and
  Toledo]{Nascimento2024_AOP_String}
F.~F. Nascimento, V.~B. Bezerra, and J.~M. Toledo.
\newblock Some remarks on hayward black hole surrounded by a cloud of strings.
\newblock \emph{Ann. Phys. (NY)}, 460:\penalty0 169548, 2024.
\newblock \doi{10.1016/j.aop.2023.169548}.

\bibitem[Perlick and Tsupko(2022)]{isz31}
V.~Perlick and O.~Yu. Tsupko.
\newblock Calculating black hole shadows: Review of analytical studies.
\newblock \emph{Phys. Rep.}, 947:\penalty0 1--39, 2022.
\newblock \doi{10.1016/j.physrep.2021.10.004}.

\bibitem[Cunha and Herdeiro(2018)]{isz32}
P.~V.~P. Cunha and C.~A.~R. Herdeiro.
\newblock Shadows and strong gravitational lensing: a brief review.
\newblock \emph{Gen. Relativ. Gravit.}, 50:\penalty0 42, 2018.
\newblock \doi{10.1007/s10714-018-2361-9}.

\bibitem[Remillard and McClintock(2006{\natexlab{a}})]{isz33}
R.~A. Remillard and J.~E. McClintock.
\newblock X-ray properties of black-hole binaries.
\newblock \emph{Annu. Rev. Astron. Astrophys.}, 44:\penalty0 49--92,
  2006{\natexlab{a}}.
\newblock \doi{10.1146/annurev.astro.44.051905.092532}.

\bibitem[Strohmayer(2001)]{isz34}
T.~E. Strohmayer.
\newblock Discovery of a 450 hz quasi-periodic oscillation from the microquasar
  gro j1655--40 with {RXTE}.
\newblock \emph{Astrophys. J. Lett.}, 552:\penalty0 L49--L53, 2001.
\newblock \doi{10.1086/320258}.

\bibitem[Stella and Vietri(1998)]{isz35}
L.~Stella and M.~Vietri.
\newblock Lense-thirring precession and quasi-periodic oscillations in low-mass
  x-ray binaries.
\newblock \emph{Astrophys. J. Lett.}, 492:\penalty0 L59--L62, 1998.
\newblock \doi{10.1086/311075}.

\bibitem[McClintock et~al.(2011)]{isz36}
J.~E. McClintock et~al.
\newblock Measuring the spins of accreting black holes.
\newblock \emph{Class. Quantum Grav.}, 28:\penalty0 114009, 2011.
\newblock \doi{10.1088/0264-9381/28/11/114009}.

\bibitem[Stuchl{\'{\i}}k et~al.(2020)Stuchl{\'{\i}}k, Kolo{\v{s}},
  Kov{\'{a}}{\v{r}}, Slan{\'{y}}, and Tursunov]{isz37}
Z.~Stuchl{\'{\i}}k, M.~Kolo{\v{s}}, J.~Kov{\'{a}}{\v{r}}, P.~Slan{\'{y}}, and
  A.~Tursunov.
\newblock Quasi-periodic oscillations as probes of strong gravity: Theoretical
  models and applications.
\newblock \emph{Universe}, 6:\penalty0 26, 2020.
\newblock \doi{10.3390/universe6020026}.

\bibitem[Konoplya and Zhidenko(2011)]{isz38}
R.~A. Konoplya and A.~Zhidenko.
\newblock Quasinormal modes of black holes: From astrophysics to string theory.
\newblock \emph{Rev. Mod. Phys.}, 83:\penalty0 793--836, 2011.
\newblock \doi{10.1103/RevModPhys.83.793}.

\bibitem[Berti et~al.(2009)Berti, Cardoso, and Starinets]{isz39}
E.~Berti, V.~Cardoso, and A.~O. Starinets.
\newblock Quasinormal modes of black holes and black branes.
\newblock \emph{Class. Quantum Grav.}, 26:\penalty0 163001, 2009.
\newblock \doi{10.1088/0264-9381/26/16/163001}.

\bibitem[Cardoso and Pani(2019)]{isz40}
V.~Cardoso and P.~Pani.
\newblock Testing the nature of dark compact objects: a status report.
\newblock \emph{Living Rev. Relativ.}, 22:\penalty0 4, 2019.
\newblock \doi{10.1007/s41114-019-0020-4}.

\bibitem[Berti et~al.(2006)Berti, Cardoso, and Will]{isz41}
E.~Berti, V.~Cardoso, and C.~M. Will.
\newblock Gravitational-wave spectroscopy of massive black holes with the space
  interferometer lisa.
\newblock \emph{Phys. Rev. D}, 73:\penalty0 064030, 2006.
\newblock \doi{10.1103/PhysRevD.73.064030}.

\bibitem[Visser(1999)]{isz42}
M.~Visser.
\newblock Some general bounds for one-dimensional scattering.
\newblock \emph{Phys. Rev. A}, 59:\penalty0 427--438, 1999.
\newblock \doi{10.1103/PhysRevA.59.427}.

\bibitem[Boonserm and Visser(2008)]{isz43}
P.~Boonserm and M.~Visser.
\newblock Bounding the bogoliubov coefficients.
\newblock \emph{Ann. Phys. (NY)}, 323:\penalty0 2779--2798, 2008.
\newblock \doi{10.1016/j.aop.2008.02.002}.

\bibitem[Synge(1960)]{JLS1960}
J.~L. Synge.
\newblock \emph{Relativity: The General Theory}.
\newblock Interscience Publishers, New York, 1960.

\bibitem[Singh et~al.(2025)Singh, Upadhyay, Myrzakulov, Myrzakulov, Singh, and
  Kumar]{DVS2025}
D.~V. Singh, S.~Upadhyay, Y.~Myrzakulov, K.~Myrzakulov, B.~Singh, and M.~Kumar.
\newblock Thermodynamic behavior and phase transitions of black holes with a
  cloud of strings and perfect fluid dark matter.
\newblock \emph{Nucl. Phys. B}, 1016:\penalty0 116915, 2025.
\newblock \doi{10.1016/j.nuclphysb.2025.116915}.

\bibitem[Wald(1984)]{sec2is08}
R.~M. Wald.
\newblock \emph{General Relativity}.
\newblock University of Chicago Press, Chicago, 1984.

\bibitem[Penrose(2002)]{sec2is09}
Roger Penrose.
\newblock ``golden oldie'': Gravitational collapse: The role of general
  relativity.
\newblock \emph{General Relativity and Gravitation}, 34:\penalty0 1141--1165,
  2002.
\newblock \doi{10.1023/A:1016578408204}.

\bibitem[Chandrasekhar(1985)]{sec4is02}
S.~Chandrasekhar.
\newblock \emph{The Mathematical Theory of Black Holes}.
\newblock Oxford University Press, Oxford, 1985.

\bibitem[Bardeen et~al.(1972)Bardeen, Press, and Teukolsky]{sec4is01}
J.~M. Bardeen, W.~H. Press, and S.~A. Teukolsky.
\newblock Rotating black holes: Locally nonrotating frames, energy extraction,
  and scalar synchrotron radiation.
\newblock \emph{Astrophys. J.}, 178:\penalty0 347--370, 1972.
\newblock \doi{10.1086/151796}.

\bibitem[Remillard and McClintock(2006{\natexlab{b}})]{sec5is01}
R.~A. Remillard and J.~E. McClintock.
\newblock X-ray properties of black-hole binaries.
\newblock \emph{Annu. Rev. Astron. Astrophys.}, 44:\penalty0 49--92,
  2006{\natexlab{b}}.
\newblock \doi{10.1146/annurev.astro.44.051905.092532}.

\bibitem[Stella and Vietri(1999)]{Stella1}
L.~Stella and M.~Vietri.
\newblock khz quasiperiodic oscillations in low-mass x-ray binaries as probes
  of general relativity in the strong-field regime.
\newblock \emph{Phys. Rev. Lett.}, 82:\penalty0 17, 1999.
\newblock \doi{10.1103/PhysRevLett.82.17}.

\bibitem[Stella et~al.(1999)Stella, Vietri, and Morsink]{Stella2}
L.~Stella, M.~Vietri, and S.~Morsink.
\newblock Correlations in the quasi-periodic oscillation frequencies of
  low-mass x-ray binaries and the relativistic precession model.
\newblock \emph{Astrophys. J. Lett.}, 524:\penalty0 L63, 1999.
\newblock \doi{10.1086/312291}.

\bibitem[Kolo{\v{s}} et~al.(2015)Kolo{\v{s}}, Stuchl{\'\i}k, and
  Tursunov]{Stuchlik}
Martin Kolo{\v{s}}, Zden{\v{e}}k Stuchl{\'\i}k, and Arman Tursunov.
\newblock {Quasi-harmonic oscillatory motion of charged particles around a
  Schwarzschild black hole immersed in a uniform magnetic field}.
\newblock \emph{Class. Quant. Grav.}, 32\penalty0 (16):\penalty0 165009, 2015.
\newblock \doi{10.1088/0264-9381/32/16/165009}.

\bibitem[Kolo{\v{s}} et~al.(2017)Kolo{\v{s}}, Tursunov, and
  Stuchl{\'\i}k]{Kolos2017}
Martin Kolo{\v{s}}, Arman Tursunov, and Zden{\v{e}}k Stuchl{\'\i}k.
\newblock {Possible signature of the magnetic fields related to quasi-periodic
  oscillations observed in microquasars}.
\newblock \emph{Eur. Phys. J. C}, 77\penalty0 (12):\penalty0 860, 2017.
\newblock \doi{10.1140/epjc/s10052-017-5431-3}.

\bibitem[Klu{\'{z}}niak and Abramowicz(2005)]{sec5is03}
W.~Klu{\'{z}}niak and M.~A. Abramowicz.
\newblock \emph{Resonant oscillations of accretion flow and KHZ QPOS}, pages
  143--148.
\newblock Springer Netherlands, Dordrecht, 2005.
\newblock ISBN 978-1-4020-4085-6.
\newblock \doi{10.1007/1-4020-4085-7_17}.
\newblock URL \url{https://doi.org/10.1007/1-4020-4085-7_17}.

\bibitem[Abramowicz and Klu{\'{z}}niak(2001)]{sec5is04}
M.~A. Abramowicz and W.~Klu{\'{z}}niak.
\newblock A precise determination of black hole spin in gro j1655--40.
\newblock \emph{Astron. Astrophys.}, 374:\penalty0 L19--L20, 2001.
\newblock \doi{10.1051/0004-6361:20010791}.

\bibitem[B{\'e}car et~al.(2024)B{\'e}car, Gonz{\'a}lez, Papantonopoulos, and
  V{\'a}squez]{Becar2023}
R.~B{\'e}car, P.~A. Gonz{\'a}lez, E.~Papantonopoulos, and Y.~V{\'a}squez.
\newblock Massive scalar field perturbations of black holes surrounded by dark
  matter.
\newblock \emph{Eur. Phys. J. C}, 84:\penalty0 329, 2024.
\newblock \doi{10.1140/epjc/s10052-024-12553-8}.

\bibitem[Liu and Zhang(2023)]{YL2023}
Y.~Liu and X.~Zhang.
\newblock Quasinormal modes of bardeen black holes with a cloud of strings.
\newblock \emph{Chin. Phys. C}, 47:\penalty0 125103, 2023.
\newblock \doi{10.1088/1674-1137/acf3d5}.

\bibitem[Boonserm(2009)]{gf4}
P.~Boonserm.
\newblock \emph{Bounding the greybody factors for black holes: A rigorous
  analytic study}.
\newblock PhD thesis, Victoria University of Wellington, 2009.

\bibitem[Boonserm et~al.(2023)Boonserm, Phalungsongsathit, Sansuk, and
  Wongjun]{gf5}
P.~Boonserm, S.~Phalungsongsathit, K.~Sansuk, and P.~Wongjun.
\newblock Greybody factors of black holes in {f(R)} gravity with matter fields.
\newblock \emph{Eur. Phys. J. C}, 83:\penalty0 657, 2023.
\newblock \doi{10.1140/epjc/s10052-023-11843-x}.

\bibitem[Sakalli and Kanzi(2022)]{gf6}
{\.I}zzet Sakalli and Sara Kanzi.
\newblock {Topical Review: greybody factors and quasinormal modes for black
  holes in various theories - fingerprints of invisibles}.
\newblock \emph{Turk. J. Phys.}, 46\penalty0 (2):\penalty0 51--103, 2022.
\newblock \doi{10.55730/1300-0101.269}.

\end{thebibliography}

\end{document}